\documentclass[aps,prc,preprint,superscriptaddress]{revtex4}

\usepackage{graphicx}
\usepackage{afterpage}
\usepackage{color}

\usepackage{natbib}

\newcommand{\Eq}[1]   {Eq.~(\ref{#1})}

\newcommand{\Fi}[1]   {Fig.~\ref{#1}}

\newcommand{\Ta}[1]   {Table~\ref{#1}}

\newcommand{\agev}    {\mbox{$A$~GeV}}               
\newcommand{\gevc}    {\mbox{GeV$/c$}}

\newcommand{\rb}[1]   {\mbox{\textrm{\scriptsize #1}}}
\newcommand{\rbt}[1]  {\mbox{\textrm{\tiny #1}}}
\newcommand{\sqrts}   {\ensuremath{\sqrt{s_{_{\rbt{NN}}}}}}

\newcommand{\lam}     {\ensuremath{\Lambda}}
\newcommand{\lab}     {\ensuremath{\bar{\Lambda}}}

\newcommand{\sig}     {\ensuremath{\Sigma^{0}}}
\newcommand{\sib}     {\ensuremath{\bar{\Sigma}^{0}}}
\newcommand{\msib}     {\ensuremath{\bar{\Sigma}^{-}}}
\newcommand{\psib}     {\ensuremath{\Sigma^{+}}}
\newcommand{\pimin}   {\ensuremath{\pi^-}}
\newcommand{\piplus}  {\ensuremath{\pi^+}}
\newcommand{\kmin}    {\ensuremath{\textrm{K}^-}}
\newcommand{\kplus}   {\ensuremath{\textrm{K}^+}}
\newcommand{\pbar}    {\ensuremath{\bar{\textrm{p}}}}

\newcommand{\pt}      {\ensuremath{p_{\rb{T}}}}
\newcommand{\mt}      {\ensuremath{m_{\rb{T}}}}

\newcommand{\mtavg}   {\ensuremath{\langle m_{\rb{T}} \rangle - m_{\rb{0}}}}

\newcommand{\dedx}    {\ensuremath{\textrm{d}E/\textrm{d}x}}
\newcommand{\dndy}    {\ensuremath{\textrm{d}n/\textrm{d}y}}

\newcommand{\nwound}  {\ensuremath{\langle N_{\rb{w}} \rangle}}
\newcommand{\sigaver}  {\ensuremath{\langle \sigma \rangle}}

\begin{document}

\title{Centrality dependence of proton and antiproton spectra in Pb+Pb collisions at 40\agev~and 158\agev~measured at the CERN SPS}




%

\affiliation{NIKHEF, 
             Amsterdam, Netherlands.}
\affiliation{Department of Physics, University of Athens, 
             Athens, Greece.}

\affiliation{E\"otv\"os Lor\'ant University, Budapest, Hungary.}

\affiliation{KFKI Research Institute for Particle and Nuclear Physics,
             Budapest, Hungary.} 
\affiliation{MIT, 
             Cambridge, USA.}
\affiliation{Henryk Niewodniczanski Institute of Nuclear Physics, 
             Polish Academy of Sciences, 
             Cracow, Poland.}
\affiliation{Gesellschaft f\"{u}r Schwerionenforschung (GSI),
             Darmstadt, Germany.} 
\affiliation{Joint Institute for Nuclear Research, 
             Dubna, Russia.}
\affiliation{Fachbereich Physik der Universit\"{a}t, 
             Frankfurt, Germany.}
\affiliation{CERN, 
             Geneva, Switzerland.}
\affiliation{Institute of Physics, Jan Kochanowski University, 
             Kielce, Poland.}
\affiliation{Fachbereich Physik der Universit\"{a}t, 
             Marburg, Germany.}
\affiliation{Max-Planck-Institut f\"{u}r Physik, 
             Munich, Germany.}
\affiliation{Charles University, Faculty of Mathematics and Physics,
             Institute of Particle and Nuclear Physics, 
             Prague, Czech Republic.} 
\affiliation{Nuclear Physics Laboratory, University of Washington,
             Seattle, WA, USA.} 
\affiliation{Atomic Physics Department, Sofia University St.~Kliment Ohridski, 
             Sofia, Bulgaria.} 
\affiliation{Institute for Nuclear Research and Nuclear Energy, 
             Sofia, Bulgaria.}
\affiliation{Department of Chemistry, Stony Brook Univ. (SUNYSB), 
             Stony Brook, USA.}
\affiliation{Institute for Nuclear Studies, 
             Warsaw, Poland.}
\affiliation{Institute for Experimental Physics, University of Warsaw,
             Warsaw, Poland.} 
\affiliation{Faculty of Physics, Warsaw University of Technology, 
             Warsaw, Poland.}
\affiliation{Rudjer Boskovic Institute, 
             Zagreb, Croatia.}


%
%

\author{T.~Anticic} 
\affiliation{Rudjer Boskovic Institute, 
             Zagreb, Croatia.}
\author{B.~Baatar}
\affiliation{Joint Institute for Nuclear Research, 
             Dubna, Russia.}
\author{D.~Barna}
\affiliation{KFKI Research Institute for Particle and Nuclear Physics,
             Budapest, Hungary.} 
\author{J.~Bartke}
\affiliation{Henryk Niewodniczanski Institute of Nuclear Physics, 
             Polish Academy of Sciences, 
             Cracow, Poland.}
\author{H.~Beck}
\affiliation{Fachbereich Physik der Universit\"{a}t, 
             Frankfurt, Germany.}
\author{L.~Betev}
\affiliation{CERN, 
             Geneva, Switzerland.}
\author{H.~Bia{\l}\-kowska} 
\affiliation{Institute for Nuclear Studies, 
             Warsaw, Poland.}
\author{C.~Blume}
\affiliation{Fachbereich Physik der Universit\"{a}t, 
             Frankfurt, Germany.}
\author{M.~Bogusz} 
\affiliation{Faculty of Physics, Warsaw University of Technology, 
             Warsaw, Poland.}
\author{B.~Boimska}
\affiliation{Institute for Nuclear Studies, 
             Warsaw, Poland.}
\author{J.~Book}
\affiliation{Fachbereich Physik der Universit\"{a}t, 
             Frankfurt, Germany.}
\author{M.~Botje}
\affiliation{NIKHEF, 
             Amsterdam, Netherlands.}
\author{P.~Bun\v{c}i\'{c}}
\affiliation{CERN, 
             Geneva, Switzerland.}
\author{T.~Cetner} 
\affiliation{Faculty of Physics, Warsaw University of Technology, 
             Warsaw, Poland.}
\author{P.~Christakoglou}
\affiliation{NIKHEF, 
             Amsterdam, Netherlands.}
\author{P.~Chung}
\affiliation{Department of Chemistry, Stony Brook Univ. (SUNYSB), 
             Stony Brook, USA.}
\author{O.~Chvala}
\affiliation{Charles University, Faculty of Mathematics and Physics,
             Institute of Particle and Nuclear Physics, 
             Prague, Czech Republic.} 
\author{J.G.~Cramer}
\affiliation{Nuclear Physics Laboratory, University of Washington,
             Seattle, WA, USA.} 
\author{V.~Eckardt}
\affiliation{Max-Planck-Institut f\"{u}r Physik, 
             Munich, Germany.}
\author{Z.~Fodor}
\affiliation{KFKI Research Institute for Particle and Nuclear Physics,
             Budapest, Hungary.} 
\author{P.~Foka}
\affiliation{Gesellschaft f\"{u}r Schwerionenforschung (GSI),
             Darmstadt, Germany.} 
\author{V.~Friese}
\affiliation{Gesellschaft f\"{u}r Schwerionenforschung (GSI),
             Darmstadt, Germany.} 
\author{M.~Ga\'zdzicki}
\affiliation{Fachbereich Physik der Universit\"{a}t, 
             Frankfurt, Germany.}
\affiliation{Institute of Physics, Jan Kochanowski University, 
             Kielce, Poland.}
\author{K.~Grebieszkow}
\affiliation{Faculty of Physics, Warsaw University of Technology, 
             Warsaw, Poland.}
\author{C.~H\"{o}hne}
\altaffiliation{now at Justus Liebig Universit\"{a}t Giessen, 35392 Giessen, Germany}
\affiliation{Gesellschaft f\"{u}r Schwerionenforschung (GSI),
             Darmstadt, Germany.} 
\author{K.~Kadija}
\affiliation{Rudjer Boskovic Institute, 
             Zagreb, Croatia.}
\author{A.~Karev}
\affiliation{CERN, 
             Geneva, Switzerland.}
\author{V.I.~Kolesnikov}
\affiliation{Joint Institute for Nuclear Research, 
             Dubna, Russia.}
\author{M.~Kowalski}
\affiliation{Henryk Niewodniczanski Institute of Nuclear Physics, 
             Polish Academy of Sciences, 
             Cracow, Poland.}
\author{D.~Kresan}
\altaffiliation{now at Justus Liebig Universit\"{a}t Giessen, 35392 Giessen, Germany} 
\affiliation{Gesellschaft f\"{u}r Schwerionenforschung (GSI),
             Darmstadt, Germany.}
\author{A.~Laszlo}
\affiliation{KFKI Research Institute for Particle and Nuclear Physics,
             Budapest, Hungary.} 
\author{R.~Lacey}
\affiliation{Department of Chemistry, Stony Brook Univ. (SUNYSB), 
             Stony Brook, USA.}
\author{M.~van~Leeuwen}
\affiliation{NIKHEF, 
             Amsterdam, Netherlands.}
\author{M.~Mackowiak} 
\affiliation{Faculty of Physics, Warsaw University of Technology, 
             Warsaw, Poland.}
\author{M.~Makariev}
\affiliation{Institute for Nuclear Research and Nuclear Energy, 
             Sofia, Bulgaria.} 
\author{A.I.~Malakhov}
\affiliation{Joint Institute for Nuclear Research, 
             Dubna, Russia.}
\author{M.~Mateev}
\affiliation{Atomic Physics Department, Sofia University St.~Kliment Ohridski, 
             Sofia, Bulgaria.} 
\author{G.L.~Melkumov}
\affiliation{Joint Institute for Nuclear Research, 
             Dubna, Russia.}
\author{M.~Mitrovski}
\affiliation{Fachbereich Physik der Universit\"{a}t, 
             Frankfurt, Germany.}
\author{S.~Mr\'owczy\'nski}
\affiliation{Institute of Physics, Jan Kochanowski University, 
             Kielce, Poland.}
\author{V.~Nicolic}
\affiliation{Rudjer Boskovic Institute, 
             Zagreb, Croatia.}
\author{G.~P\'{a}lla}
\affiliation{KFKI Research Institute for Particle and Nuclear Physics, Budapest, Hungary.} 
\author{A.D.~Panagiotou}
\affiliation{Department of Physics, University of Athens, 
             Athens, Greece.}
\author{W.~Peryt}
\affiliation{Faculty of Physics, Warsaw University of Technology, 
             Warsaw, Poland.}
\author{J.~Pluta}
\affiliation{Faculty of Physics, Warsaw University of Technology, 
             Warsaw, Poland.}
\author{D.~Prindle}
\affiliation{Nuclear Physics Laboratory, University of Washington,
             Seattle, WA, USA.} 
\author{F.~P\"{u}hlhofer}
\affiliation{Fachbereich Physik der Universit\"{a}t, 
             Marburg, Germany.}
\author{R.~Renfordt}
\affiliation{Fachbereich Physik der Universit\"{a}t, 
             Frankfurt, Germany.}
\author{C.~Roland}
\affiliation{MIT, 
             Cambridge, USA.}
\author{G.~Roland}
\affiliation{MIT, 
             Cambridge, USA.}
\author{M.~Rybczy\'nski}
\affiliation{Institute of Physics, Jan Kochanowski University, 
             Kielce, Poland.}
\author{A.~Rybicki}
\affiliation{Henryk Niewodniczanski Institute of Nuclear Physics, 
             Polish Academy of Sciences, 
             Cracow, Poland.}
\author{A.~Sandoval}
\affiliation{Gesellschaft f\"{u}r Schwerionenforschung (GSI),
             Darmstadt, Germany.} 
\author{N.~Schmitz}
\affiliation{Max-Planck-Institut f\"{u}r Physik, 
             Munich, Germany.}
\author{T.~Schuster}
\affiliation{Fachbereich Physik der Universit\"{a}t, 
             Frankfurt, Germany.}
\author{P.~Seyboth}
\affiliation{Max-Planck-Institut f\"{u}r Physik, 
             Munich, Germany.}
\author{F.~Sikl\'{e}r}
\affiliation{KFKI Research Institute for Particle and Nuclear Physics, Budapest, Hungary.} 
\author{E.~Skrzypczak}
\affiliation{Institute for Experimental Physics, University of Warsaw,
             Warsaw, Poland.} 
\author{M.~Slodkowski}
\affiliation{Faculty of Physics, Warsaw University of Technology, 
             Warsaw, Poland.}
\author{G.~Stefanek}
\affiliation{Institute of Physics, Jan Kochanowski University, 
             Kielce, Poland.}
\author{R.~Stock}
\affiliation{Fachbereich Physik der Universit\"{a}t, 
             Frankfurt, Germany.}
\author{H.~Str\"{o}bele}
\affiliation{Fachbereich Physik der Universit\"{a}t, 
             Frankfurt, Germany.}
\author{T.~Susa}
\affiliation{Rudjer Boskovic Institute, 
             Zagreb, Croatia.}
\author{M.~Szuba}
\affiliation{Faculty of Physics, Warsaw University of Technology, 
             Warsaw, Poland.}
\author{M.~Utvi\'{c}}
\affiliation{Fachbereich Physik der Universit\"{a}t, 
             Frankfurt, Germany.}
\author{D.~Varga}
\affiliation{KFKI Research Institute for Particle and Nuclear Physics, Budapest, Hungary.} 
\affiliation{CERN, 
             Geneva, Switzerland.}
\author{M.~Vassiliou}
\affiliation{Department of Physics, University of Athens, 
             Athens, Greece.}
\author{G.I.~Veres}
\affiliation{KFKI Research Institute for Particle and Nuclear Physics, Budapest, Hungary.} 
\affiliation{MIT, Cambridge, USA.}
\author{G.~Vesztergombi}
\affiliation{KFKI Research Institute for Particle and Nuclear Physics, Budapest, Hungary.}
\author{D.~Vrani\'{c}}
\affiliation{Gesellschaft f\"{u}r Schwerionenforschung (GSI),
             Darmstadt, Germany.} 
\author{Z.~W{\l}odarczyk}
\affiliation{Institute of Physics, Jan Kochanowski University, 
             Kielce, Poland.}
\author{A.~Wojtaszek-Szwarc}
\affiliation{Institute of Physics, Jan Kochanowski University, 
             Kielce, Poland.}





\date{\today }

\begin{abstract}
The yields of (anti-)protons were measured by the NA49 Collaboration in centrality selected Pb+Pb collisions at 40\agev~and 158\agev. Particle identification was obtained in the laboratory momentum range from 5 to 63 \gevc~by the measurement of the energy loss \dedx~in the TPC detector gas. The corresponding rapidity coverage extends 1.6 units from mid-rapidity into the forward hemisphere.
Transverse mass spectra, the rapidity dependences of the average transverse mass, and rapidity density distributions were studied as a function of collision centrality. 
The values of the average transverse mass as well as the midrapidity yields of protons when normalized to the number of wounded nucleons show only modest centrality dependences. In contrast, the shape of the rapidity distribution changes significantly with collision centrality, especially at 40\agev. The experimental results are compared to calculations of the HSD and UrQMD transport models.
\end{abstract}



\maketitle
\setlength{\topmargin}{0.1cm}
\section{Introduction}
It is generally accepted that heavy ion collisions at ultra-relativistic energies result in a 
fireball of  matter with high density and temperature. Such conditions prevail, when the 
incoming nucleons deposit enough of their kinetic energy in the reaction zone. Little is known
about this stopping process and about possible differences between the stopping of the incident 
nucleons in elementary nucleon-nucleon interactions and nucleus-nucleus collisions. In a simple
microscopic picture each nucleon interacts only once in elementary 
interactions, whereas in central nuclear collisions multiple interactions prevail. There are three 
methods to study nuclear stopping. In proton-nucleus (p+A) collisions the difference between the 
c.m.-energies of the incident and the most forward going nucleon is a good measure of the stopped 
energy. This type of analysis has been pioneered in reference~\cite{Busza}. Another approach to 
study effects due to multiple projectile nucleon collisions employs symmetric collisions of nuclei of
different size. The resulting distributions of participating nucleons in terms of 
longitudinal and transverse momenta may vary with the size of the incident nuclei. Such 
differences reflect the change of stopping power as a function of system size. Finally, the 
effective system size can be changed by varying the impact parameter in collisions between heavy 
nuclei. Here again the momentum distribution of participating nucleons may be different in central and 
peripheral collisions as a consequence of changes in the stopping behavior.

 Energy loss in central collisions was studied as a function of beam energy at AGS in Brookhaven~\cite{Back:2000ru}, at CERN SPS in Geneva~\cite{Appelshauser:1998yb} and at RHIC in Brookhaven~\cite{Bearden:2003hx}. It was found that the rapidity distribution changes from a convex to a concave form from AGS to RHIC via SPS energies. At AGS energies the energy loss was
 also studied as a function of collision centrality~\cite{Back:2000ru}. A strong centrality dependence was observed. This topic has 
 also been addressed by NA49 in two previous publications. The first one presents the net-proton  distribution for central Pb+Pb collisions at 158\agev~\cite{Appelshauser:1998yb}. The second one describes the measurement of midrapidity proton and antiproton yields in Pb+Pb collisions
at various energies from 20$A$ to 158\agev~as well as for different centralities at 158\agev~\cite{Alt:2005gr}. At RHIC the PHOBOS collaboration has published the centrality dependence of the rapidity density at midrapidity and the transverse mass spectra of net protons \cite{phobos1} at \sqrts~=~62.4 GeV and charged particle pseudorapidity distributions \cite{phobos2,phobos3} at 
\sqrts~=~19.6~GeV, 62.4~GeV, and 130~GeV. Data on centrality dependent particle production at $y$~=~0 and $y$~=~1 in Au+Au collisions at \sqrts~=~200~GeV \cite{Brahms} are available from the BRAHMS collaboration. The STAR collaboration has published Au+Au data on the centrality dependence of proton and antiproton production for $\left|y\right|<$ 0.5 at \sqrts~=~130~GeV \cite{Star}.

In this paper we present the centrality dependence of proton and antiproton transverse mass and rapidity  distributions in Pb+Pb collisions at 40\agev~and 158\agev~as obtained with the NA49 detector~\cite{Milica}.
The phase space coverage extends in center-of-mass rapidity $y$ from mid-rapidity 1.6 units into the forward hemisphere and ranges from zero to 2.0 \gevc~in transverse momentum \pt. Net-proton distributions at 158\agev~are obtained from those of all protons by subtracting the distributions of antiprotons. This 
analysis supplements the NA49 data on $\pi$, K, $\phi$, $\lam$ production as function of beam energy 
at the CERN SPS~\cite{na49_piK,na49_phi,na49_hyp}. Pion and kaon production in Pb+Pb collisions at 40\agev~and 158\agev~is analyzed as function of centrality with the same method and is the subject of a separate paper~\cite{na49_piK_cent}.

\section{Experimental setup and data sets}

\begin{table}[!h]
	\centering
\caption{Cross section fractions 
in \% centrality, average numbers of wounded nucleons \nwound, and numbers of analyzed events  for the five centrality classes at
40\agev~and 158\agev. Only systematic errors are quoted.}
\vspace{\baselineskip}
\begin{tabular} {|c||c|c|c|c|}
\hline
centrality class& centrality [ \% ] & \nwound~& analyzed events \\
\hline
\multicolumn {4}{|c|} {40\agev~}\\
\hline\hline
C0 & 0-5 &351 $\pm$ 3& 13034\\
C1 & 5-12.5 &290 $\pm$ 4 & 22971\\
C2 & 12.5-23.5 & 210 $\pm$ 6 & 34035\\
C3 & 23.5-33.5&142 $\pm$ 8 &32668\\
C4 & 33.5-43.5& 93 $\pm$ 7 & 32071\\
\hline\hline
\multicolumn {4}{|c|} {158\agev~}\\
\hline\hline
C0 & 0-5 &352 $\pm$ 3& 15306\\
C1 & 5-12.5 &281 $\pm$ 4 & 23548\\
C2 & 12.5-23.5 & 196 $\pm$ 6 & 37053\\
C3 & 23.5-33.5&128 $\pm$ 8 &34554\\
C4 & 33.5-43.5& 85 $\pm$ 7 & 34583\\
\hline
\end{tabular}
\label{tab:cent}
\end{table}

The NA49 detector is a large acceptance hadron spectrometer at the CERN SPS~\cite{NA49NIM}. The main components 
are four large time projection chambers (TPCs) and two super-conducting dipole magnets with a 1 m 
vertical gap, aligned in a row, and a total bending power of 9~Tm. 
Two 2 m long TPCs (VTPCs) inside the magnets each with 72 pad-rows along the beam direction allow for precise tracking, momentum determination, 
vertex reconstruction, and particle identification (PID) by the measurement of the energy 
loss (\dedx) in the detector gas. The other two TPCs (MTPCs)) have large dimensions (4m x 4m x 1.2m, 90 pad-rows) and provide 
additional momentum resolution for high momentum particles as well as PID by \dedx~measurement 
with a resolution of around 4\%.
Two time-of-flight scintillator arrays of 891 pixels each,
situated just behind the MTPCs symmetrically on either side
of the beam axis, supplement particle identification in the
momentum range from 1 to 10 GeV/c.
A Veto Calorimeter (VCAL), which is placed further downstream
along the beam  and covers the projectile spectator phase space region, is used  to select event 
centrality. The NA49 detector is described in detail in reference~\cite{NA49NIM}.
The Pb beam had a typical intensity of 10$^{4}$ ions/s and impinged on a target Pb foil
with a (areal) density of 224 mg/cm$^2$.  It passed through a quartz Cherenkov detector from which the start signal for the time-of-flight measurement was obtained, and three stations of multi-wire proportional chambers which measured the trajectories of individual beam particles. A minimum bias trigger was derived from the signal of a gas Cherenkov device right behind the
target. Only interactions which reduce the beam charge and thus the signal seen by 
this detector by at least 10\% are accepted. The interaction cross 
section thus defined is 5.7 b. The contamination by background events remaining after cuts 
on vertex position and quality amounts to less than 5\% for the most peripheral collisions and is negligible for near-central collisions (see~\cite{ALaszlo}). The resulting event ensemble  
was divided into five centrality classes C0, C1, C2, C3, and C4 
(see \Ta{tab:cent} and~\cite{ALaszlo}). The centrality 
selection is based on the forward going energy of projectile spectators as measured in VCAL. The 
number of projectile spectator nucleons and the average number of interacting (wounded) nucleons 
\nwound\  were calculated from the selected cross section fractions using the Glauber approach. 
The track finding efficiency and \dedx\  resolution were optimized by track quality criteria. To 
be accepted, a track must have at least 50 (out of a maximum of 90) potential points in the MTPCs and 
have at least 5 measured and 10 potential points in one of the VTPCs.
Finally, tracks were required to have azimuthal angles within +/- 30 degrees with respect to the bending plane in order to minimize reconstruction inefficiencies and to optimize the accuracy of the \dedx~measurements.

\section{Analysis method}

\begin{figure}[h]
\includegraphics[width=0.4\linewidth]{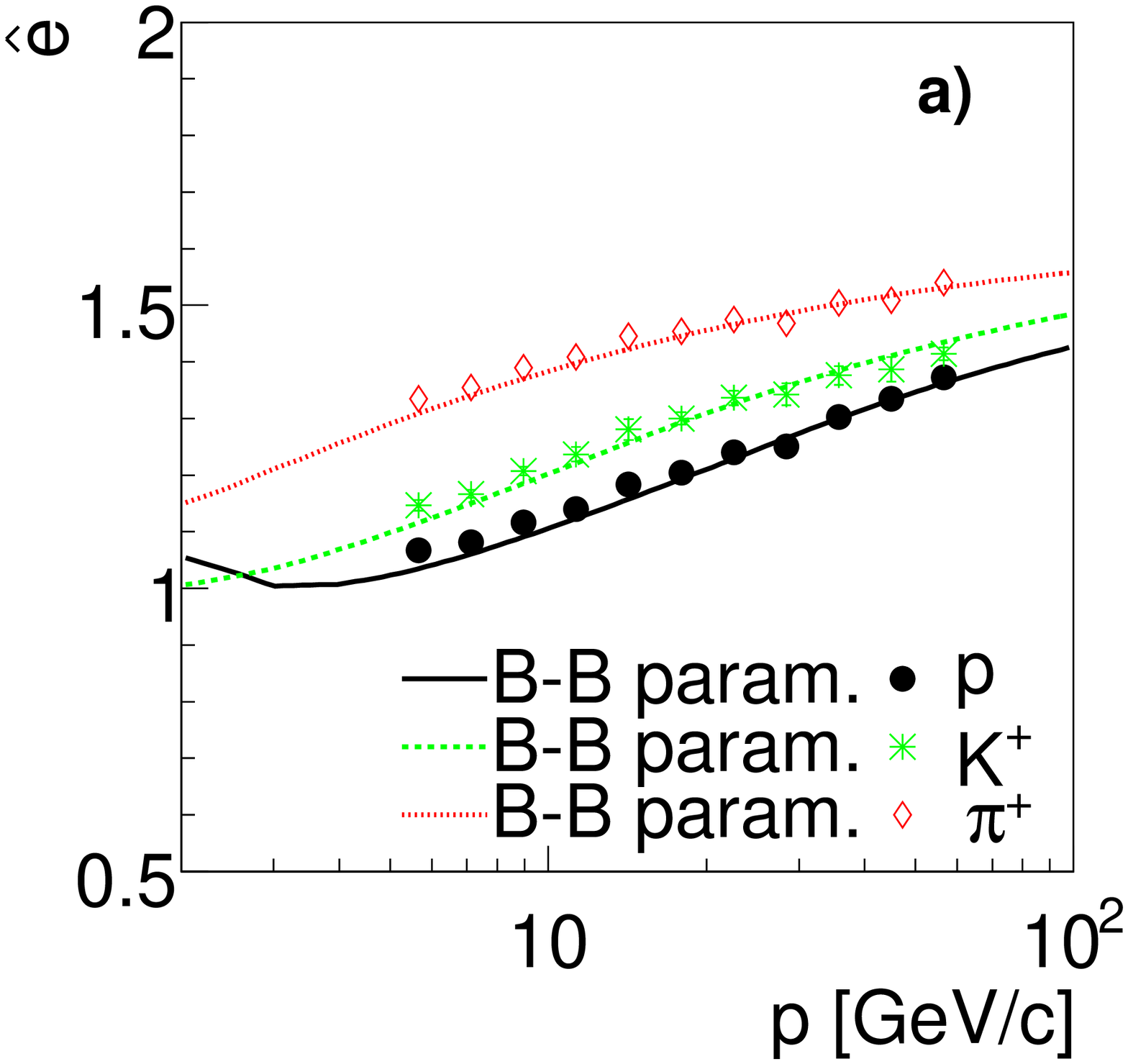}
\includegraphics[width=0.4\linewidth]{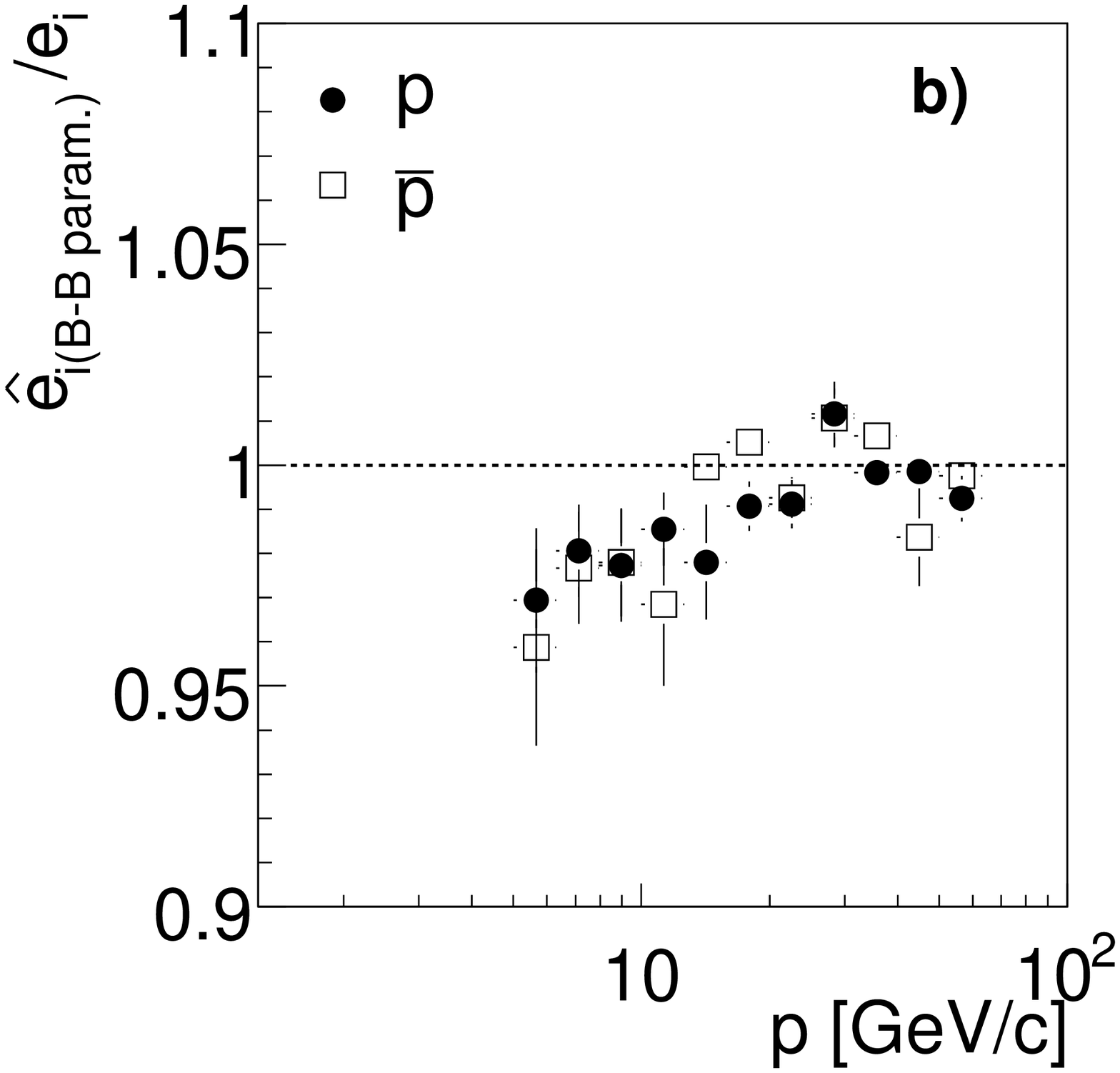}
\caption{a) The most probable specific energy loss $\hat{e}$
for particle species $\pi^{+}$, K$^{+}$ und p as a function of total momentum $p$. The points show results extracted from data whereas the lines indicate the Bethe-Bloch parametrizations optimized for the NA49 measurement. b) The ratios of the Bethe-Bloch parametrizations ($e_i$) to  $\hat{e}_i$ from data are shown as function of laboratory momentum. The index $i$ stands for p and \pbar.
}
\label{fig:comp_pos_bb} 
\end{figure}

The protons are identified by the measurement of their specific 
energy loss in the relativistic rise region. As an appropriate measure of \dedx~we calculated for each track the truncated mean of the distribution of charges measured in each pad row of the MTPCs.   
Their raw yields were extracted by fitting the function $F$
to the \dedx~distribution of all positively charged particles in 
narrow bins of total momentum $p$ and transverse momentum \pt~\cite{MvL}.
Antiproton yields are determined from the \dedx~distributions of all negatively charged particles with the same method at 158\agev. 
At 40\agev~the antiproton statistics was very low and did not allow reliable 
extraction of yields. 
The shape of $F$ is assumed to be the sum of Gaussians. 
Their parameters depend on the particle masses and the measured track lengths.
We modified the Gaussian functions by means of an extra asymmetry parameter
to account for tails of the Landau distributions which are still present 
even after truncation.

The function $F$ reads

\begin{eqnarray}\label{dedx-fit}
F{\left(\frac{\mathrm{d}E}{\mathrm{d}x}\right)}
 = \sum_{i=d,p,K,\pi,e} A_{i} \frac{1}{\sum_{l}n_{l}}\sum_{l}\frac{n_{l}}{\sqrt{2\pi}\sigma_{i,l}}
\exp 
\left[
-\frac{1}{2}{\left(\frac{e_i(p) - \hat{e}_i(p)}
{(1\pm\delta)\sigma_{i,l}}
\right)}^2
\right].
\end{eqnarray}

Here $\frac{\mathrm{d}E}{\mathrm{d}x}$ (abreviated $e$) is the measure of specific ionization, and $\hat{e}$ 
is its most probable value. The other parameters of the function are:
\begin{itemize}
	\item $A_{i}$: the raw yield of the particle $i$ under consideration in a given phase space bin.
	\item $n_{l}$: the number of tracks in a given track length interval $l$. The second sum together 
	with the normalization ${\sum_{l}n_{l}}$ forms the weighted average of the track ensembles in each phase space interval.
	\item $\sigma_{i,l}$: the width for the asymmetric Gaussian of particle type $i$ in length interval $l$.
	\item $\delta$: the asymmetry parameter.
\end{itemize}

\begin{figure}[h]
\includegraphics[width=0.3\linewidth]{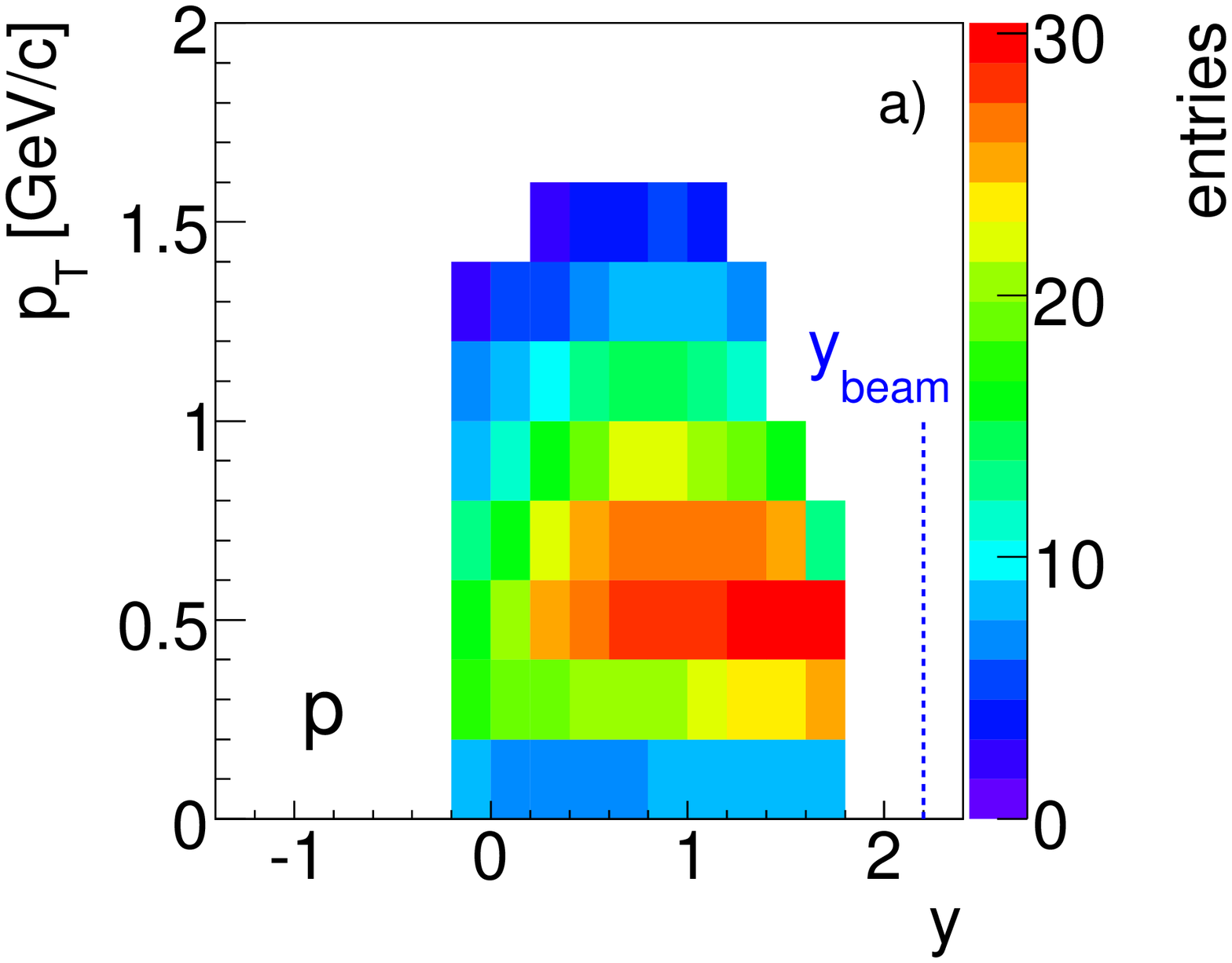}
\includegraphics[width=0.3\linewidth]{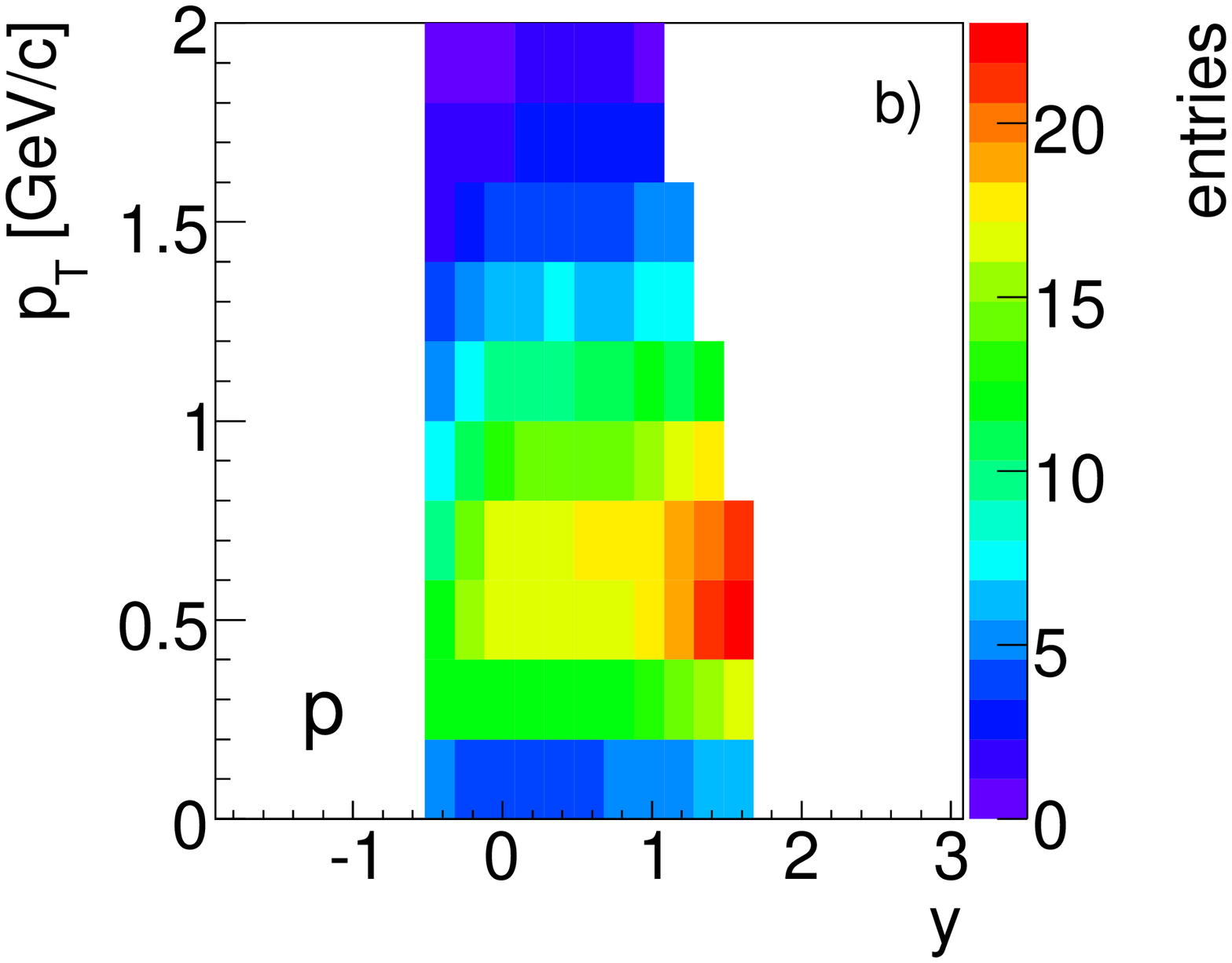}
\includegraphics[width=0.3\linewidth]{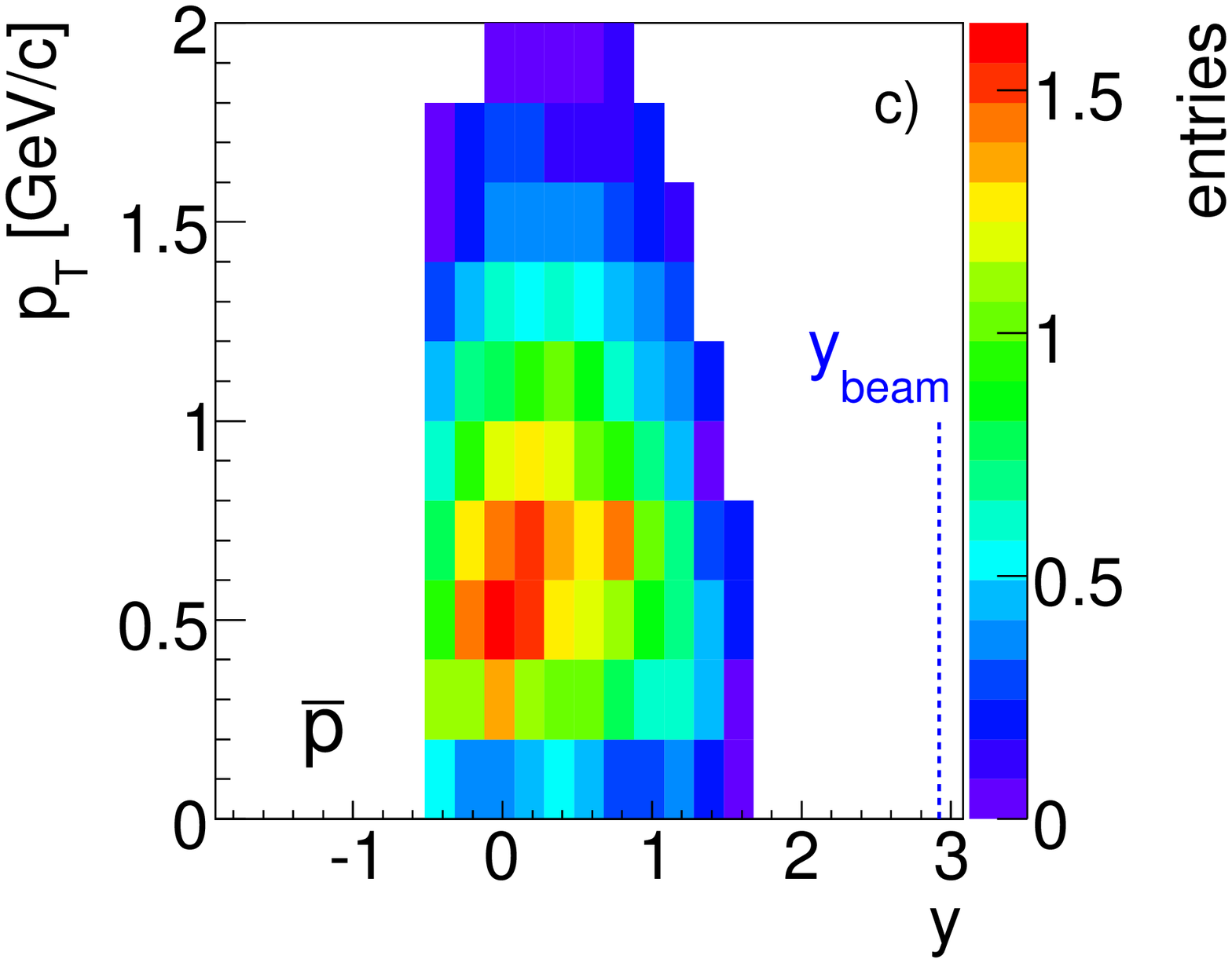}
\caption{Raw yields of protons at 40\agev~(a), protons (b) and antiprotons (c) at 158\agev~are shown as function of center-of-mass rapidity $y$ and \pt~for centrality class C2.}
\label{fig:raw_yields} 
\end{figure}

The amplitude parameters $A_{i}(p,\pt)$ were determined by a maximum likelihood fit to the \dedx~histogram in each $p$, \pt\ bin. The peak positions $\hat{e_{i}}(p)$ are considered to be \pt\ independent and were determined by fits to the 
\pt\ integrated distributions in all $p$ bins. The fitted peak positions $\hat{e_{i}}(p)$ for $\pi^{+}$, K$^{+}$ and p are compared to a parametrization of the momentum dependence of the most probable energy loss determined for the NA49 detector in \Fi{fig:comp_pos_bb}a. The relative differences between fitted (anti-)proton positions and the Bethe-Bloch parameterization are smaller than 5\% (see \Fi{fig:comp_pos_bb}b) . We have verified that the peak positions do not depend on centrality.
The widths of the Gaussians ($\sigma_{i,l}$) depend on the particle type $i$ and on the track length $l$ according to $\sigma_{i,l} = \sigma\cdot\large(\hat{e_{i}}/{e_{\pi}\large)^\alpha} (1 / \sqrt{l}) $. The exponent $\alpha$ (= 0.625) was extracted from simultaneous fits to $m^2$ distributions from TOF and to \dedx\ distributions from the TPCs~\cite{MvL}. The momentum averaged widths \sigaver\ were determined for each centrality bin by fits to the data in the whole $p$ range. These momentum averaged widths \sigaver\  turned out to be approximately 4 \% for each centrality bin.
$\delta$ was studied by fits with the asymmetric Gaussians. Since
it did not show any significant variation with centrality, total and transverse momentum~\cite{Milica}, this parameter was fixed to a constant value (0.071). The total number of fit parameters for each $p$, \pt~bin is 9. 
The raw yields were transformed from a fine grid in $\log(p)$, \pt\ to a coarser grid in $y$,~\pt. Examples of the transformed results for centrality class C2 are shown in \Fi{fig:raw_yields}.

\subsection{Acceptance and inefficiency}
The raw particle yields have to be corrected for losses due to tracks which do not pass through the detectors or which do not fulfill the
 acceptance criteria (acceptance losses) and tracks which are not properly reconstructed (efficiency losses).  
The acceptance was calculated by generating a sample of (anti-)protons with flat distributions in 
transverse momentum and rapidity. The generated particles were propagated through 
the detectors (and the magnetic field) using the programs provided in the GEANT 3.21~\cite{geant} package. 
Along the resulting trajectories realistic detector signals were generated and processed in exactly 
the same way as the experimental data. The ratio of generated to accepted particle tracks in each 
 $y$,~\pt\ bin is the acceptance correction factor.
In addition to the well-defined corrections necessary to correct for the limited and
constrained acceptance, the raw spectra may be subject to losses due to detector occupancy and thus centrality dependent inefficiencies. These losses
were minimized by restricting the analysis to tracks with azimuthal emission angles within $\pm$ 30 degrees with respect to the bending plane and in the bending direction. The remaining losses were determined by the
following procedure: Ten GEANT generated (anti-)proton tracks and
their signals were embedded into raw data of real events. Only those tracks are embedded which pass all acceptance
criteria. These modified events are reconstructed with the standard
reconstruction chain. The ratio of all generated tracks to those
reconstructed constitutes the applied correction factor for reconstruction 
inefficiencies. The resulting efficiencies vary with centrality by less than 5\% and are over 95\% in most $y$,~\pt\ bins. 
In the further analysis the bin size in rapidity of the antiproton spectra (see \Fi{fig:raw_yields}c) was doubled in order to reduce the statistical errors. The one in \pt~was also doubled at 40\agev.

\subsection{Feed-down corrections}
The measured (anti-)proton yield contains (anti-)protons from weak
hyperon decays, namely of the \lam~(\lab), the \sig~(\sib), and the \psib~(\msib). The (anti-)proton contribution from \psib~(\msib) decays is determined by scaling the estimated
feed-down correction coming from \lam~(\lab) by the ratio \lam~/ \psib~(\lab~/\msib) derived from  
a statistical hadron gas model~\cite{Becattini:2005xt} thus assuming the same phase space distributions for \lam~(\lab) and \psib~(\msib) hyperons. To determine the 
feed-down correction $n_{\rb{fd}}$ from $\lam$ ($\lab$) in each measured phase space bin,
 we used a similar procedure as for the efficiency calculation. 
\lam~(\lab) were generated according to distributions
measured by NA49~\cite{na49_hyp,na49_hyp_cent} and embedded into real events. Those (anti-)protons 
from embedded \lam~(\lab) decays,  which are reconstructed and accepted by the track selection cuts as such, have to be subtracted from the 
(anti-)proton yield after proper normalization per event (see \Eq{fd}). In reference~\cite{na49_hyp,na49_hyp_cent} 
the  measured \lam~(\lab) include the  \lam~(\lab) from electromagnetic decays of \sig~ und \sib~. The feed-down correction is given by:
\begin{eqnarray}
\label{fd}
n_{fd}(y, \pt) = \frac{N^{found}_{\lam~(\lab)}(y,\pt) }
         {N^{sim}_{\lam~(\lab)}(y,\pt)}  Y_{\rb{tot}}(y, \pt)
\end{eqnarray}
where $N^{found}_{\lam~(\lab)}(y,\pt)$ is the average number of reconstructed (anti-)protons per event from embedded \lam~(\lab) decays, $N^{sim}_{\lam~(\lab)}(y,\pt) $ is the number of simulated \lam~(\lab) and $Y_{\rb{tot}}(y,\pt) $ the multiplicity of \lam~(\lab) in the phase space interval from reference~\cite{na49_hyp,na49_hyp_cent} scaled for the contributions from \psib~(\msib). 
The lowest and the two highest rapidity bins in the 40\agev~data (see \Fi{fig:raw_yields}a) were removed due to low statistics in too many \pt~bins. 

\subsection{Systematic Errors}

One of the sources for systematic errors is the uncertainty in the procedure of unfolding the \dedx~distributions. Studies of the 
sensitivity to small changes of the fit parameters showed that the systematic errors are of the same order as the statistical errors. The largest contribution to the systematic error was traced to the deviations of the peak positions from the predicted values (see \Fi{fig:comp_pos_bb}). They reach up to 4\% in
the yields in the low momentum bins. The next source of systematic errors is the uncertainty related to acceptance, inefficiencies and feed-down correction calculations. To estimate the magnitude of these errors cut parameters were varied such that significantly different correction factors were obtained. The resulting final results varied only within the statistical errors. Overall it appears that the systematic uncertainties resulting from the correction procedures are of the order of 3\% or less. 

The determination of the rapidity density distributions required extrapolations of the transverse momentum distributions. Their contribution is negligible for most of the rapidity bins except the rapidity bins $y~\geq$~1.2, where the extrapolation factors reached values of 1.3.
Single and double exponentials were used to describe the shape of the transverse momentum distributions in this region. The resulting differences of $p_T$-integrated yields are smaller than 2\%. Finally, the results from the \dedx\ analysis in this paper and the TOF-results~\cite{Alt:2005gr} agree within 5 \% in the common acceptance  region at all centralities. We conclude from these studies that each data point carries a systematic uncertainty of approximately 7 \%.

\section{Results and comparison with models}\label{results}

\subsection{Transverse momentum spectra}
\begin{figure}[h]
\includegraphics[width=0.32\linewidth]{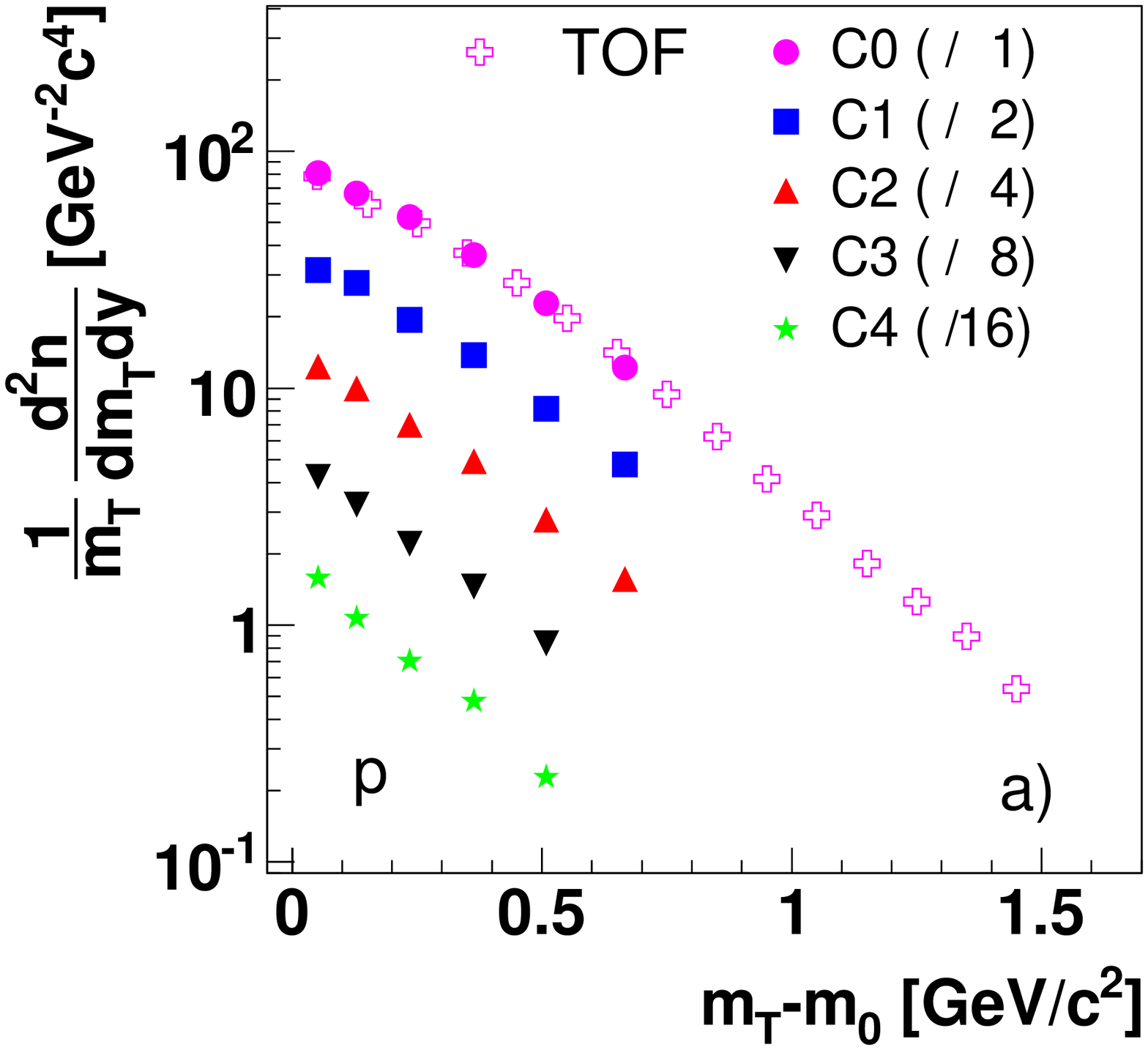}
\includegraphics[width=0.32\linewidth]{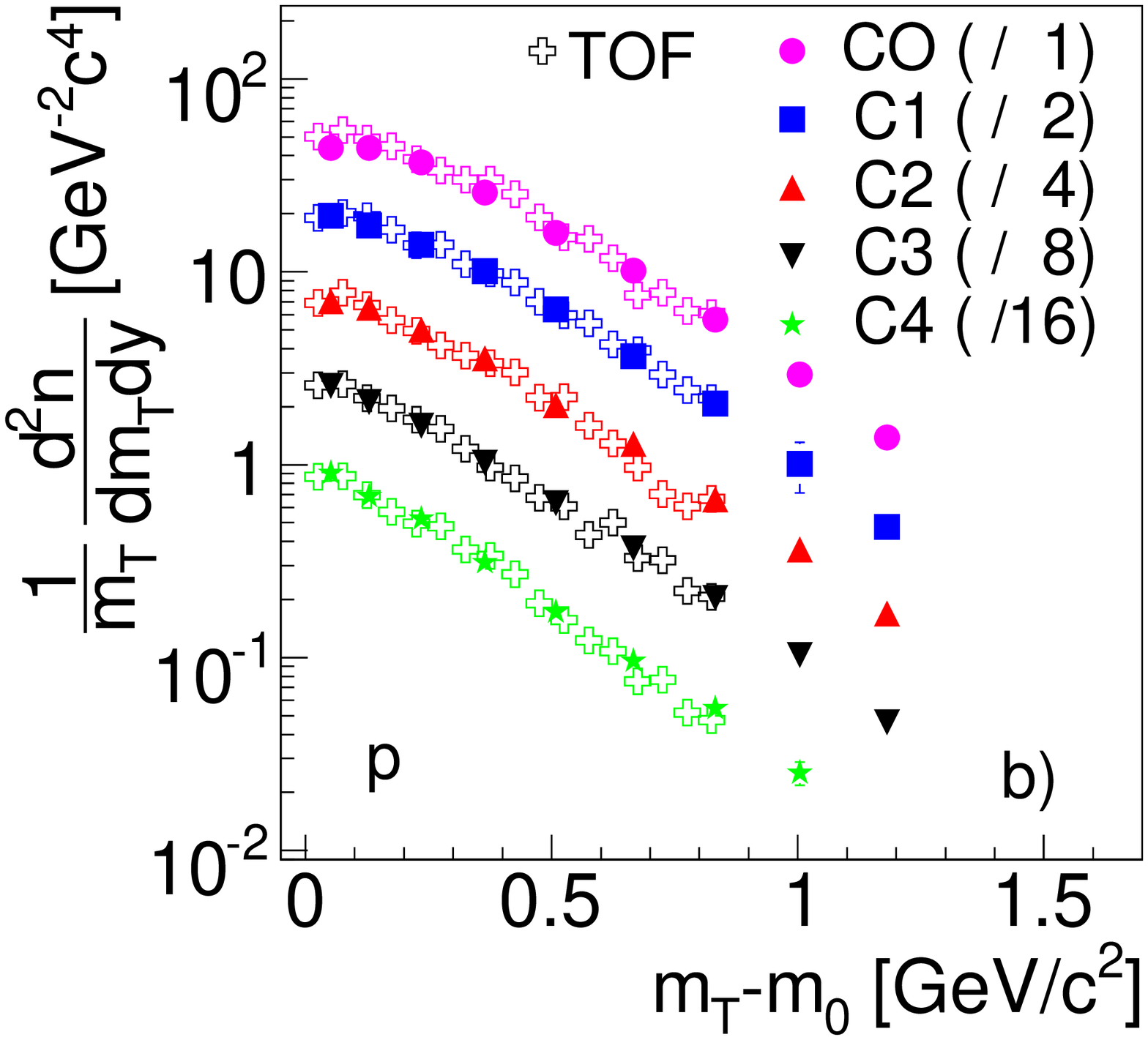}
\includegraphics[width=0.32\linewidth]{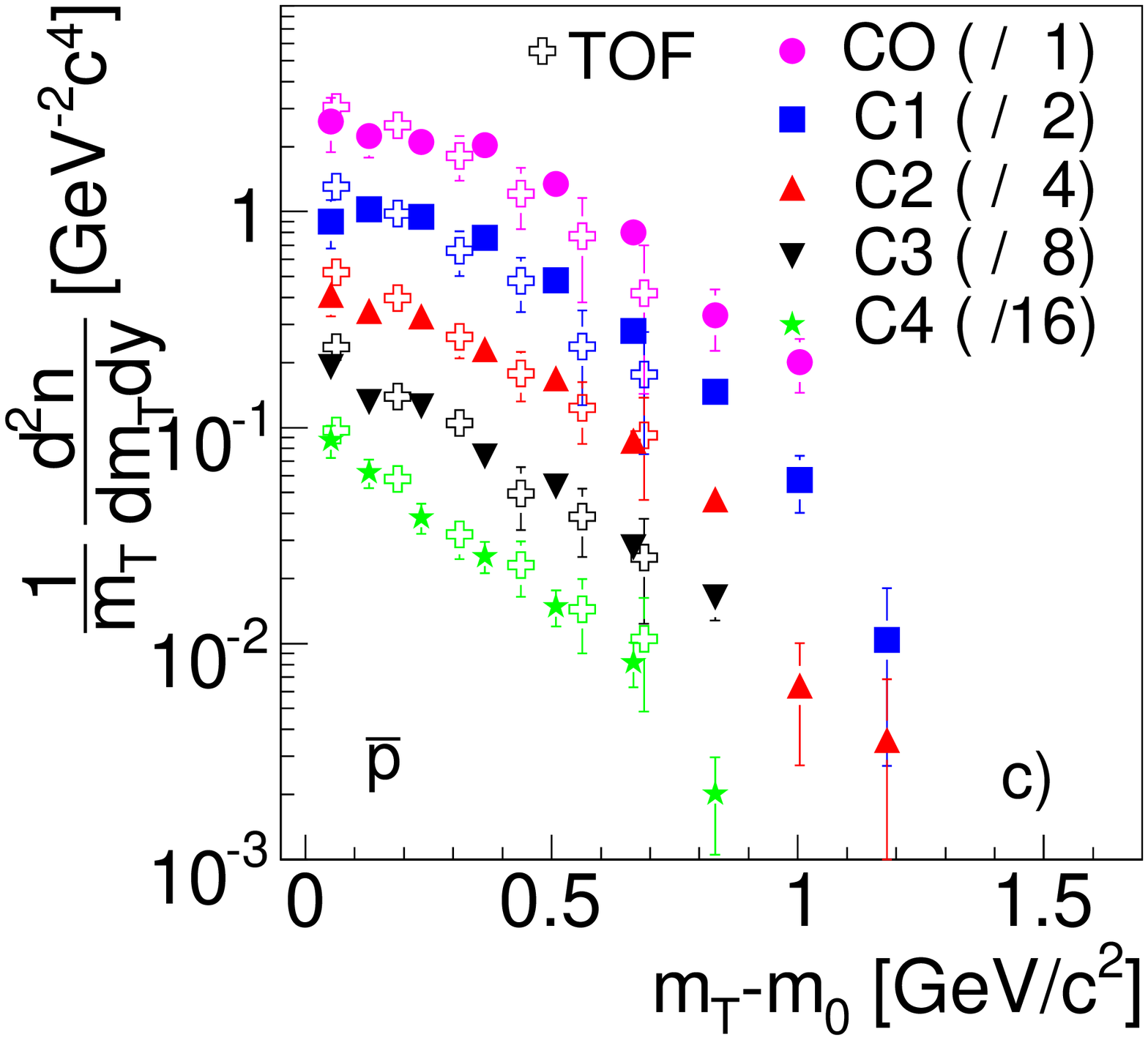}
\caption{Invariant \mt~spectra of protons at 40\agev~(a) and 158\agev~(b) as well as antiprotons at 158\agev~(c) for five centrality bins in Pb+Pb collisions. The respective center-of-mass rapidity intervals are $~-0.02~<~y~<~0.18, ~-0.12~<~y<~0.08, ~-0.12~<~y ~<~0.08$. The data at different centralities are scaled down by the factors indicated in the figures. The new NA49 measurements (full symbols) are compared, whenever available, to results of an earlier analysis  using TOF information for particle identification {\protect\cite{Alt:2005gr} (open symbols)}.
Only statistical errors are shown if larger than the symbol size. For the systematic errors see section IIIC.}
\label{fig:comp_mt_tof}
\end{figure} 

\begin{figure}[hb]
\includegraphics[width=0.95\linewidth]{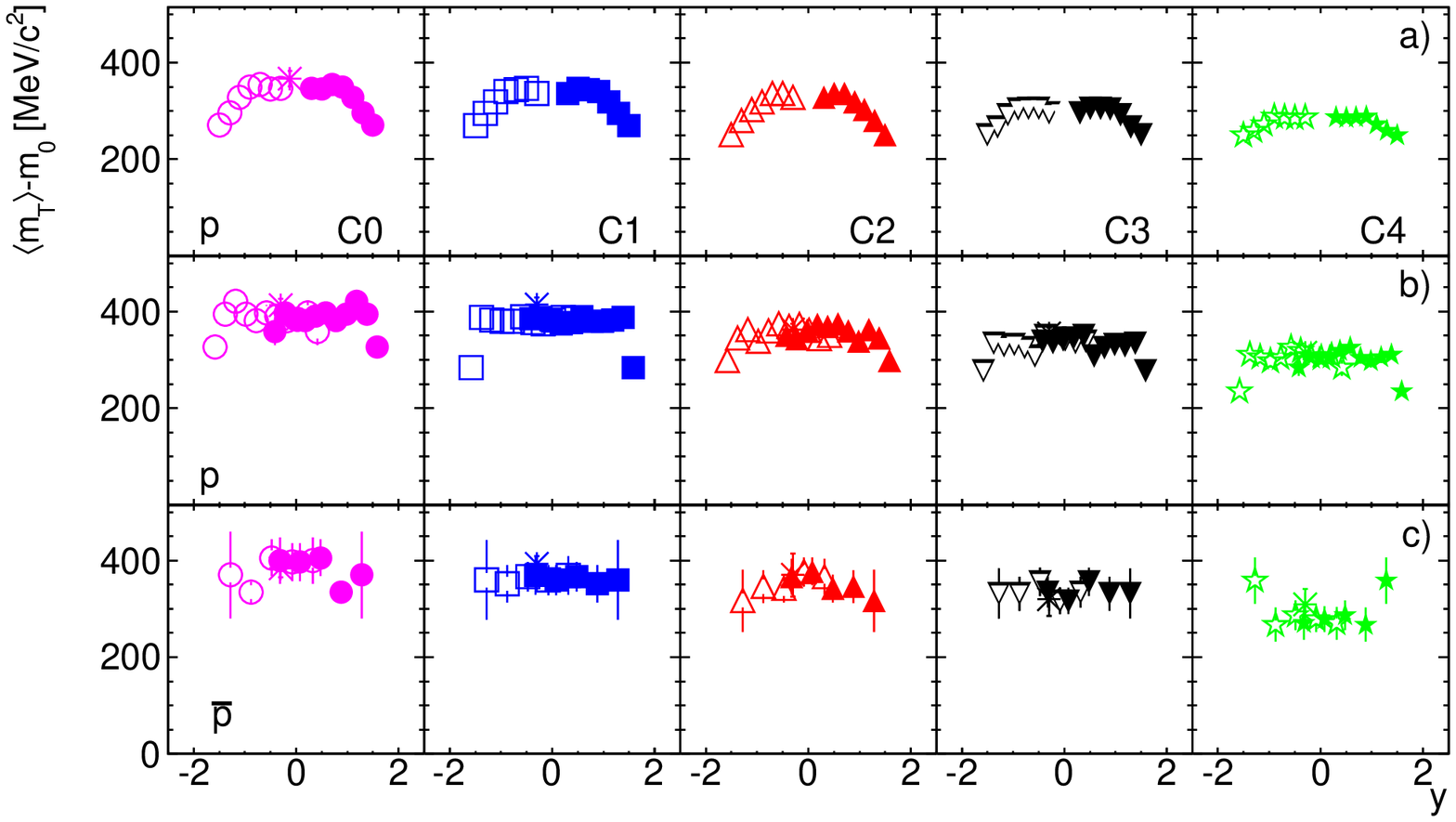}
\caption{\mtavg~for (anti-)protons as function of rapidity for five centrality intervals in Pb+Pb collisions. We show in the upper row (a)
protons at 40\agev~, in the middle row (b) protons at 158\agev, and in the lower row (c) antiprotons at 158\agev. The open symbols are obtained by reflection at mid-rapidity. Asterisks refer to results of an earlier analysis using TOF information for particle identification {\protect\cite{Alt:2005gr}}.Only statistical errors are shown (if larger than the symbol size). For the systematic errors see section IIIC.}
\label{fig:mean_mt} 
\end{figure} 

Transverse momentum distributions of (anti-)protons were determined in 10 (11) bins of rapidity at 40\agev~(158\agev)~as shown in \Fi{fig:raw_yields}. The range covered in transverse momentum extends from \pt~equal zero to 0.8~\gevc~at high rapidity and up to 1.5~\gevc~(2~\gevc) for $0.0 <y < 1.0 $ at 40\agev~(158\agev). Whenever necessary the \pt~spectra were extrapolated to 2~\gevc~(neglecting contributions at higher transverse momenta which have been accounted for in the systematic uncertainties) 
by taking the mean of the single and double exponential fits to the measured invariant spectra. At 40\agev~the extrapolation of the two bins around midrapidity were performed with the functional form obtained from the adjacent rapidity bin because of low statistics in the \pt~distributions.
Using theses extrapolations $p_T$ integrated yields (\dndy) and mean \mt~values (\mtavg) as function of rapidity were calculated (see below). Midrapidity invariant \mt~spectra $(1/{\mt~}~{{\textrm{d}}^{2}n}/{({\textrm{d}}\mt\ {\textrm{d}}y)})$ at 40\agev~and 158\agev~are compared in \Fi{fig:comp_mt_tof} with the results of a TOF based analysis published previously~\cite{Alt:2005gr}. We find agreement within errors except
for some of the two lowest \mt~ points with deviations up to 20\% in the 158\agev~ data. These differences can be traced to the feed-down corrections, which relied on different (older) parameterizations used for the MC input of (anti-)lambda phase space distributions in reference~\cite{Alt:2005gr}, but are based on recent measured data in this analysis (see above). However, this uncertainty has only a small effect on the value of \dndy.

\begin{figure}[h]
\includegraphics[width=0.32\linewidth]{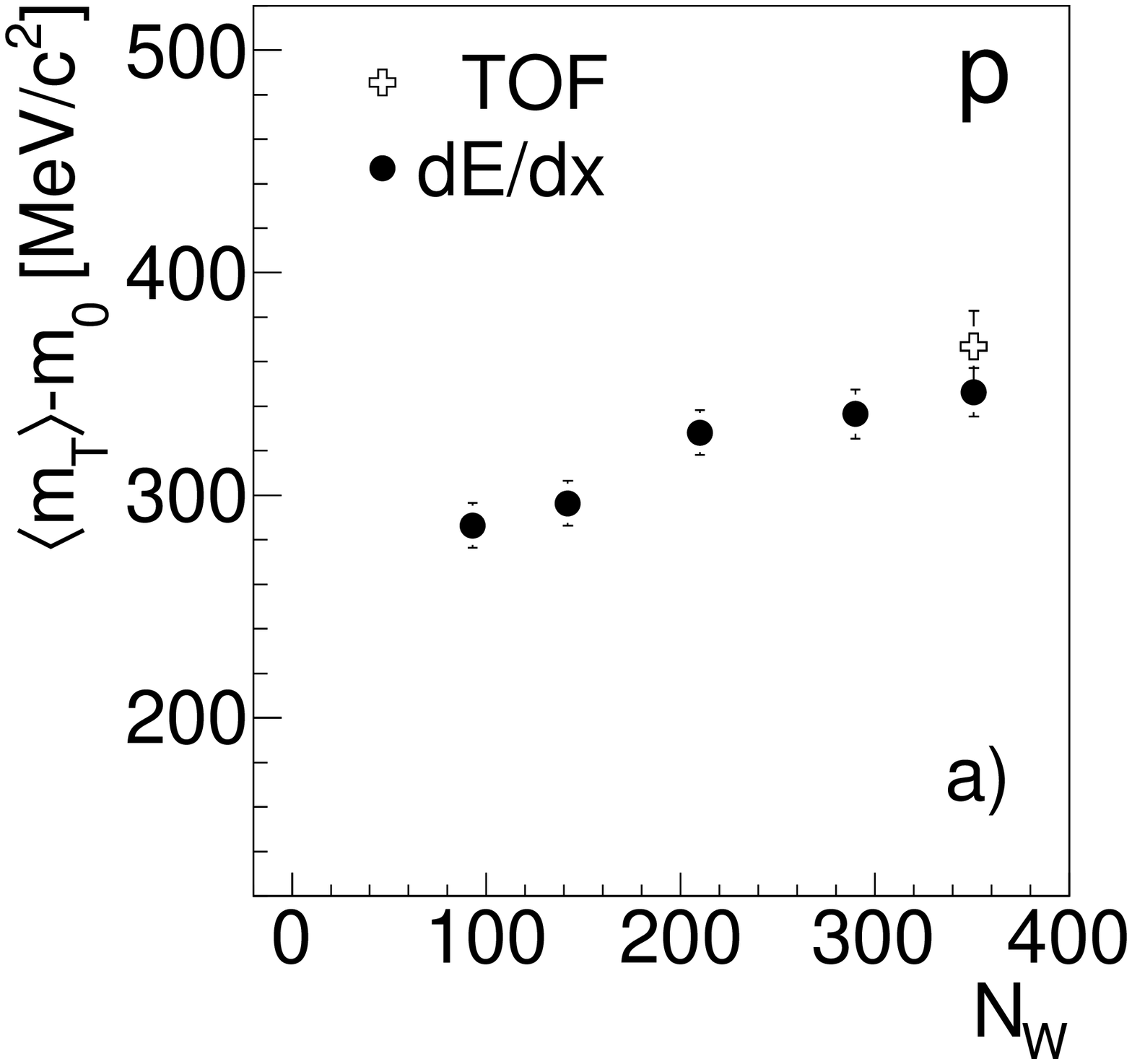}
\includegraphics[width=0.32\linewidth]{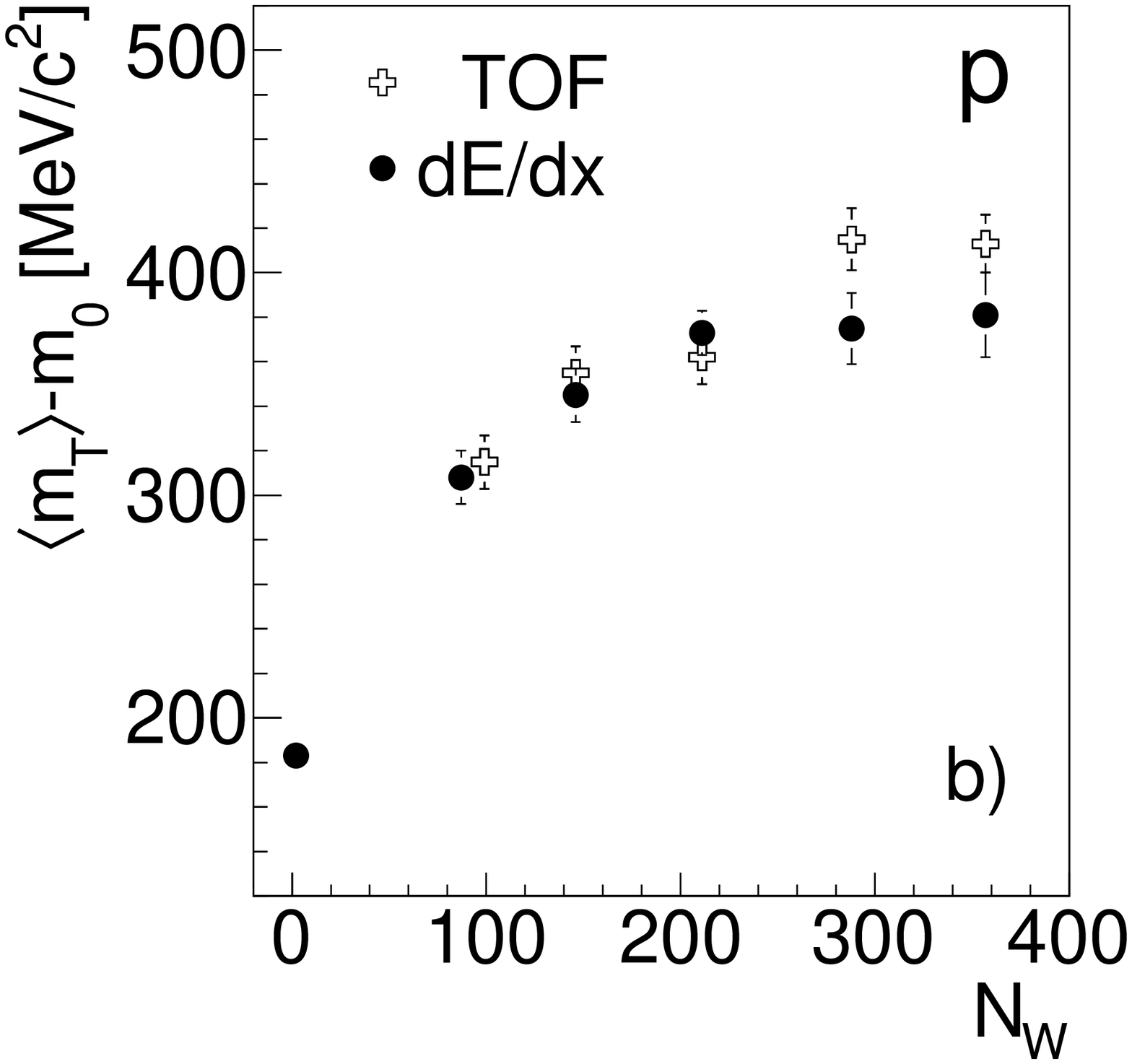}
\includegraphics[width=0.32\linewidth]{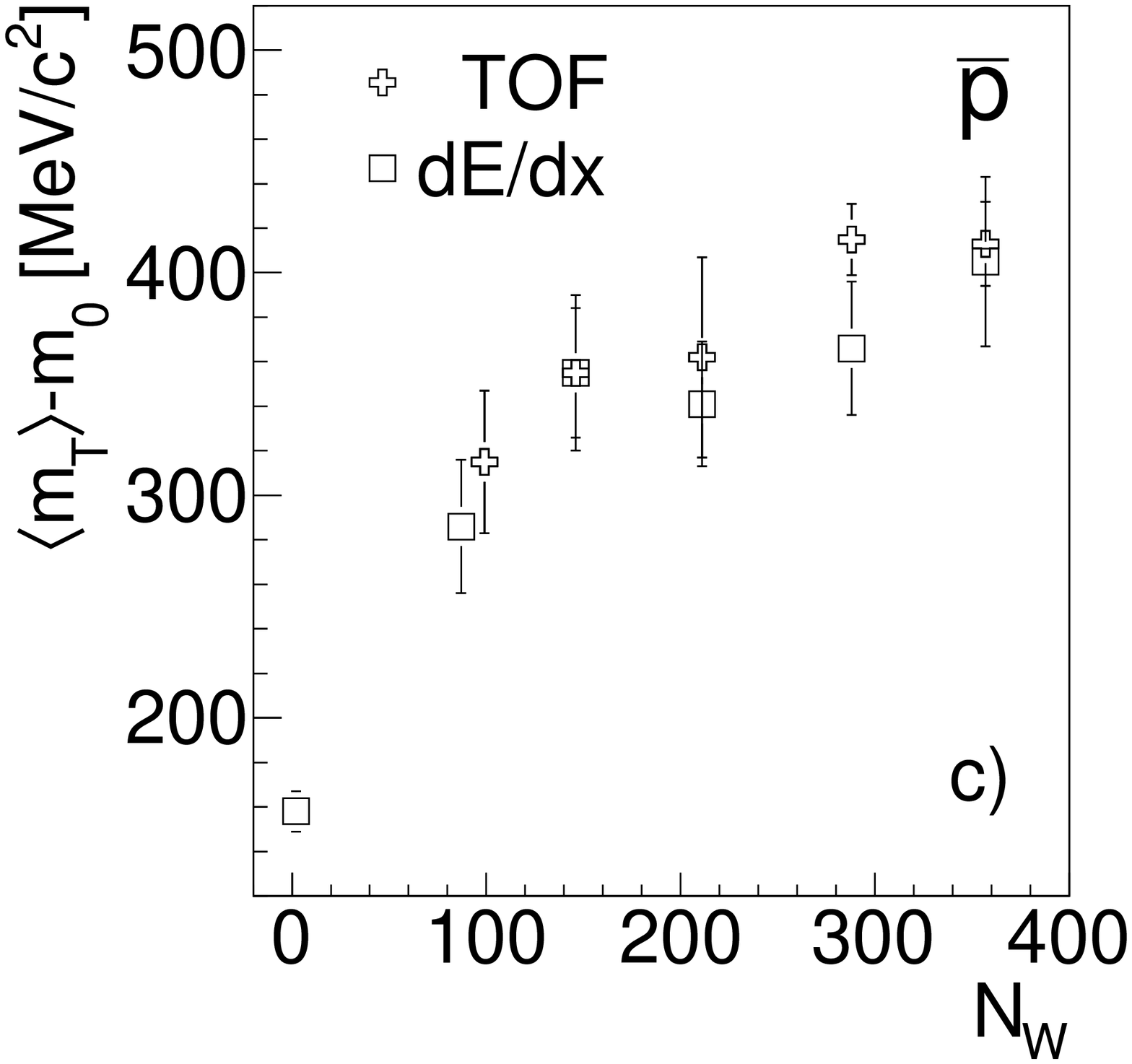}
\caption{\mtavg~ near mid-rapidity as function of \nwound~ in Pb+Pb collisions. The leftmost panel (a) shows the proton data at 40\agev, the middle panel (b) the proton data at 158\agev, and the right panel (c) the antiproton data at 158\agev. The results of this analysis (labelled \dedx) are compared to previously published Pb+Pb data (labelled TOF) {\protect\cite{Alt:2005gr}}. The ordinate has a suppressed zero. Only statistical errors are shown (if larger than the symbol size). For the systematic errors see section IIIC.}
\label{fig:mean_mt_y0} 
\end{figure} 
Since the shapes of all invariant \mt\ spectra deviate significantly from single exponentials, we choose \mtavg\ instead of the inverse slope parameter of the transverse mass spectra to study the transverse activity as function of rapidity. \Fi{fig:mean_mt} shows \mtavg\ of protons at 40\agev~(upper row) and 158\agev~(middle row), and antiprotons at 158\agev~(lower row) as function of cms rapidity. In the latter the bin size in rapidity was doubled to reduce the statistical errors on the data points. The \mtavg~values near midrapidity are plotted in \Fi{fig:mean_mt_y0} as function of centrality. Also shown are the data points from the TOF analysis~\cite{Alt:2005gr} and recent results from p+p interactions at the same energy ~\cite{Fischer}. A clear increase of \mtavg\ by roughly 30\% is observed when comparing the values obtained in the most peripheral with the most central event sample. The increase is close to a factor of two when the results on p+p interactions are taken as reference. Similar observations have been made by NA49 for data on hyperons~\cite{na49_hyp_cent}.

\subsection{Rapidity spectra}

\begin{figure}[h]
\includegraphics[width=0.48\linewidth]{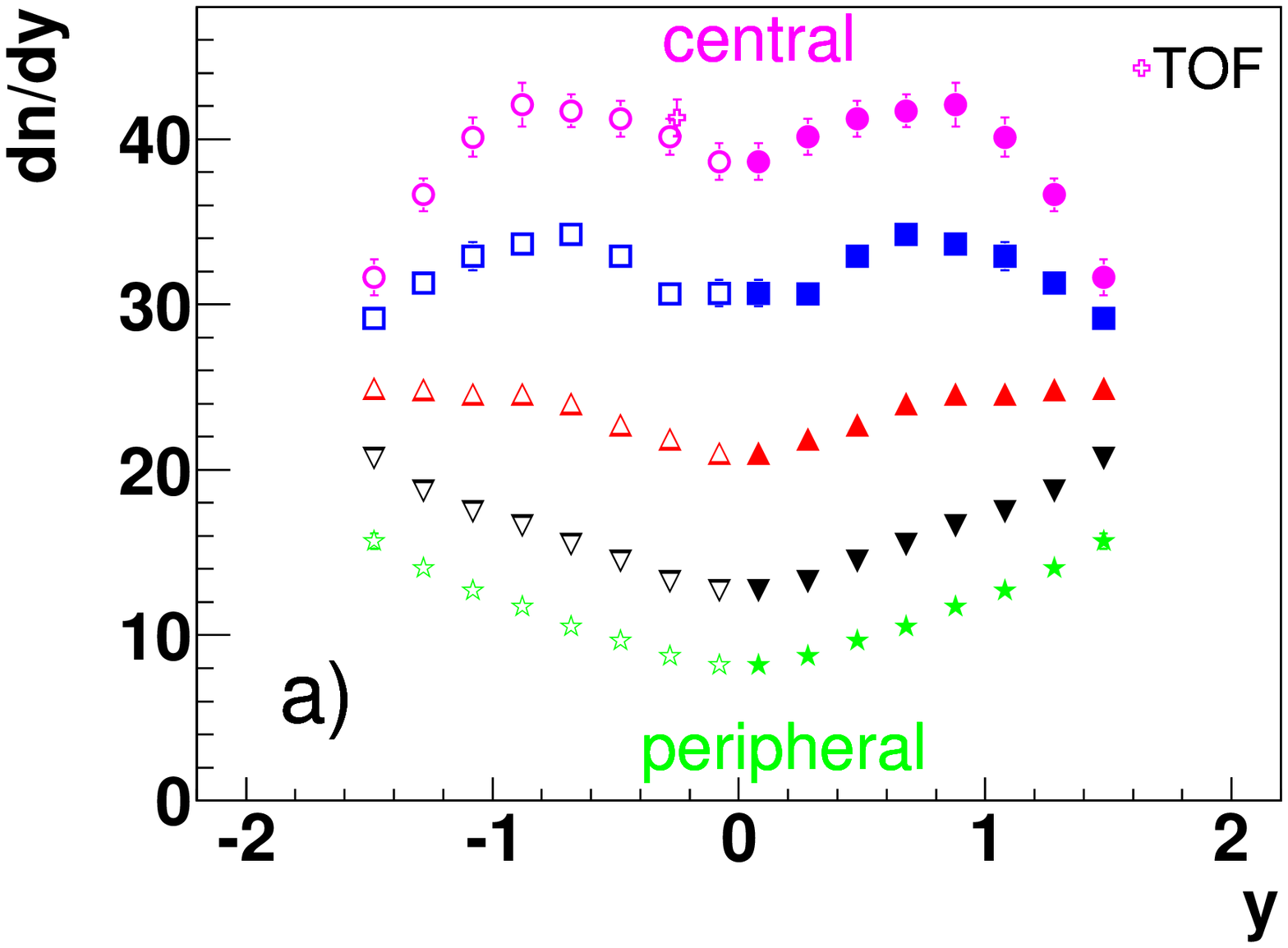}
\includegraphics[width=0.48\linewidth]{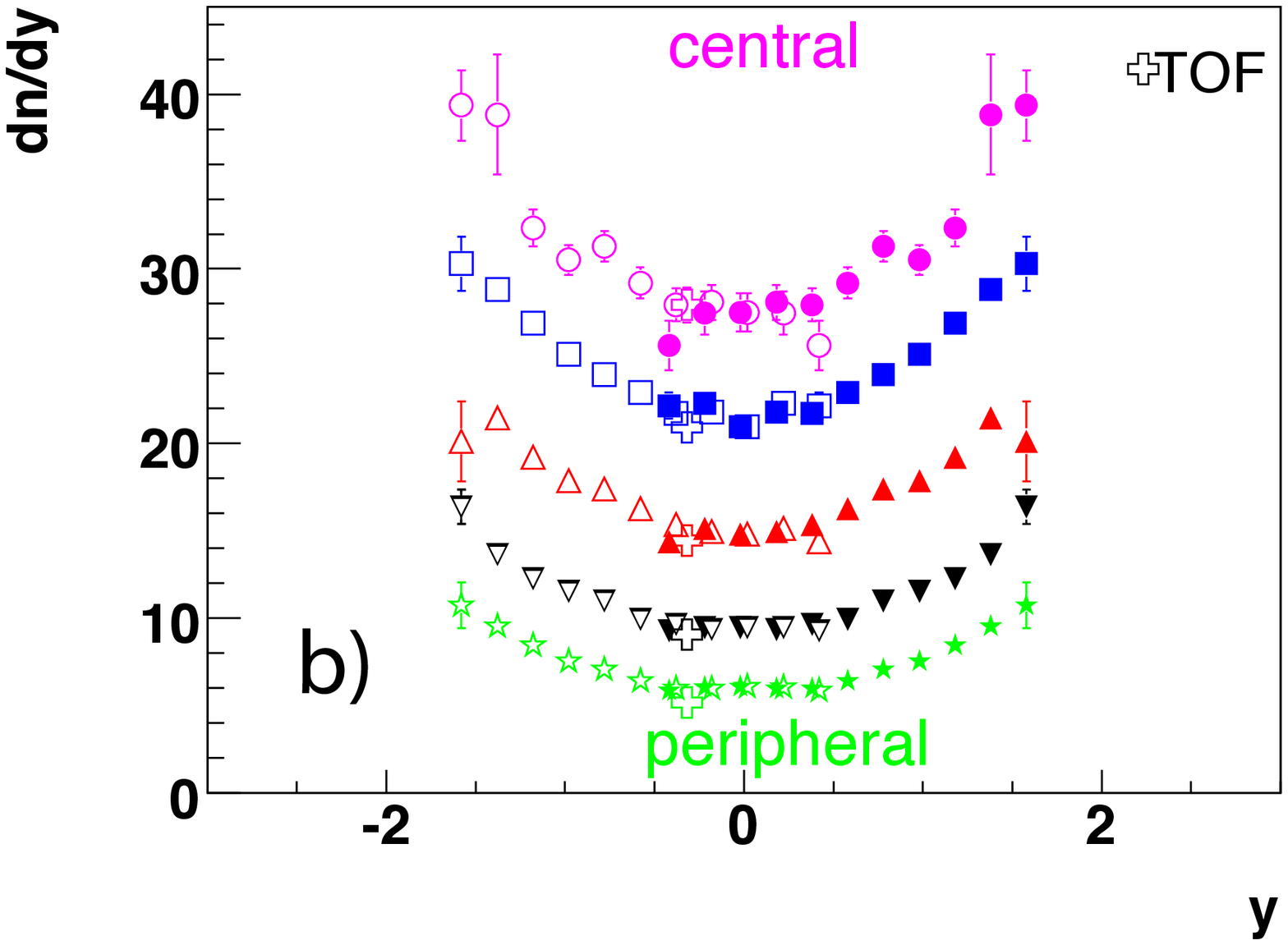}
\caption{$dn/dy$ of (net) protons for five different centralities. We show in the left panel (a) the proton distributions at 40\agev~and in the right panel (b) the difference between proton and antiproton spectra (i.e. the net proton spectra) at 158\agev. The open symbols are obtained by reflection at mid-rapidity. Also shown are the results from an earlier analysis labelled TOF {\protect\cite{Alt:2005gr}}. Only statistical errors are shown (if larger than the symbol size). For the systematic errors see section IIIC. }
\label{fig:dndy_p_40_158} 
\end{figure} 

\begin{figure}[h]
\includegraphics[width=0.99\linewidth]{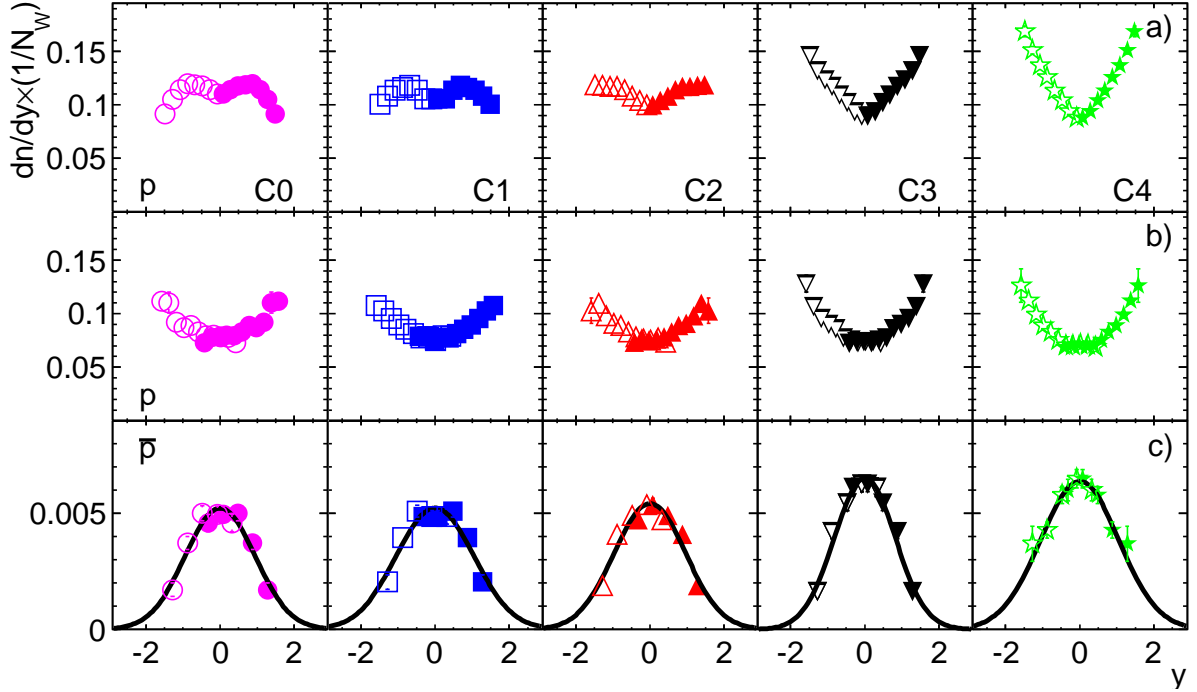}
\caption{Rapidity density distributions per wounded nucleon are shown for protons at 40\agev~(top panel), protons at 
158\agev~(middle panel), and antiprotons at 158\agev~(bottom panel) for five centralities in Pb+Pb collisions.
The open symbols are obtained by reflection at mid-rapidity. The solid lines represent results of single Gaussian fits. For the statistical errors see \Fi{fig:dndy_p_40_158}. For the systematic errors see section IIIC. } 
\label{fig:cc_dndy_all_p_ap} 
\end{figure}

\begin{figure}[h]
\includegraphics[width=0.65\linewidth]{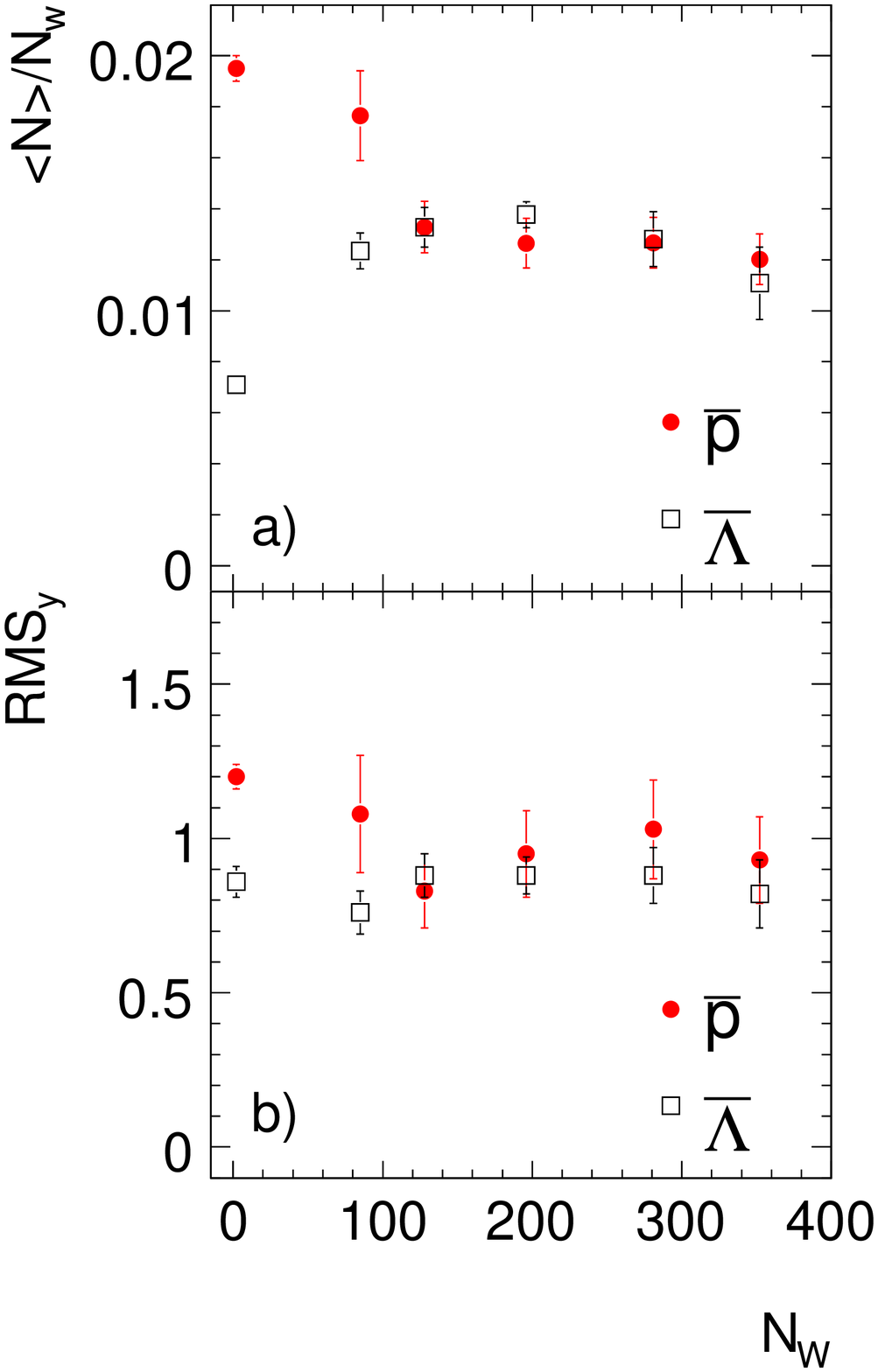}
\caption{The total multiplicities (per wounded nucleon) of \pbar~ and \lab~(upper panel) and the RMS$_{y}$ widths of the \dndy~distributions of \pbar~ and \lab~(lower panel) for five centralities as function of the number of wounded nucleons \nwound~ in Pb+Pb collisions at 158\agev. The \lab~ data are from {\protect\cite{na49_hyp_cent}}. Also shown are results on antiprotons obtained from p+p interactions {\protect\cite{Fischer}}. Only statistical errors are shown (if larger than the symbol size). For the systematic errors see section IIIC. }
\label{fig:tot_mult_pbar_2} 
\end{figure} 

The rapidity densities \dndy~in each rapidity interval were
obtained by summing the yields in the measured \pt\ interval and the integrals of the extrapolation function above the highest measured 
\pt~bin. The additive correction was calculated as the average of the single and the double exponentials in the unmeasured \pt~region.
The contributions of the extrapolation to \dndy\ are mostly of order of 3~\%  (6~\%) for protons and below 5~\% for antiprotons at 158\agev~(40\agev) and contribute negligibly to the errors of the integrals. These extrapolations reach up to 25 \% close to the limits of the accepted rapidity region.

The rapidity spectra of protons at 40\agev~(left) and net protons at 158\agev~(right) are presented in \Fi{fig:dndy_p_40_158} for five different centrality selections (C0 - C4). We included the \dndy\ values obtained from the TOF analysis (asterisks) published earlier by NA49~\cite{Alt:2005gr} for comparison. At both energies the yields increase with centrality. No change in shape is apparent at 158\agev~beam energy, whereas at 40\agev~the form of the 
\dndy\ distribution evolves from a parabolic shape near to midrapidity ($|y|~<~1$) in semi-peripheral to a double hump structure in central collisions. The \dndy~values of protons at 40\agev~are given in \Ta{tab:prot_40}, those for protons at 158\agev~in \Ta{tab:prot_158} and for antiprotons in \Ta{tab:aprot_158}.

The trends in the evolution of the rapidity distributions are seen best, when the spectra are divided by the number of wounded nucleons as shown in \Fi{fig:cc_dndy_all_p_ap}. The proton spectra at 
158\agev~(middle panel) change little. The normalized yields (at midrapidity) increase by roughly 15\% from 0.065 to 0.075 when going from semi-peripheral to central collisions.
At 40\agev~the scaled yields at midrapidity increase by 25\% from 0.085 to 0.11 from semi-peripheral to central collisions. As to the shape it seems that with decreasing centrality 
additional protons populate the region $|y|~<~(y_{max}-1.2)$. The shape of the antiproton distributions at 158\agev~resembles a Gaussian and does not change with centrality. The integrated and mid-rapidity yields  normalized by \nwound~decrease by 20\% with increasing centrality. 

We present the dependences on centrality of the normalized antiproton multiplicity and of the width of their rapidity distribution
in \Fi{fig:tot_mult_pbar_2} together with
NA49 data from elementary p+p interactions~\cite{Fischer} both at 
158\agev. The mean multiplicities (total yield per event) of antiprotons for different centralities were calculated by integrating the measured rapidity spectra and by extrapolating into the unmeasured regions assuming a Gaussian shape (see \Fi{fig:cc_dndy_all_p_ap}). The magnitude of the corresponding extrapolation factors are in the range from 5\% to 10 \%. A double Gaussian fit is used to estimate the systematic errors of the extrapolation into the high $y$-region.
The antiproton multiplicity normalized to \nwound, shown in the upper panel of \Fi{fig:tot_mult_pbar_2}, increases by nearly a factor 1.5 when going from mid-central Pb+Pb to inelastic p+p collisions and stays constant from mid-central to central Pb+Pb collisions. 
Although the properly weighted average of yields in p+p and n+n collisions 
would be the appropriate reference for this comparison, we consider the p+p midrapidity yield to be a good approximation. It is interesting to note that the antilambda multiplicity (from reference~\cite{na49_hyp_cent}), also shown in \Fi{fig:tot_mult_pbar_2}a, exhibits a centrality dependence similar to the one of the antiprotons.
The lower panel of \Fi{fig:tot_mult_pbar_2} shows the centrality dependence of the widths of the rapidity distributions for antiprotons and antilambdas (from reference~\cite{na49_hyp_cent}). No significant difference is observed between the values in different centrality bins of the Pb+Pb data. Only the width in p+p interactions seems to be slightly wider than those observed in Pb+Pb collisions. 

Net-proton rapidity distributions for all five centralities are obtained by 
subtracting the antiproton distributions from those of the protons.
The result at 158\agev~for central collisions (bin C0) is compared to 
data published earlier~\cite{Appelshauser:1998yb} in 
\Fi{fig:C0_net-prot_comp}. The differences between the two measurements 
can be traced back to different analysis methods. The earlier analysis
used a method in which distributions of negatively charged particles were subtracted from those of the positively charged ones assuming the proton mass for all particles. The resulting distributions were corrected for the then unmeasured differences between \piplus~ and \pimin, as well as \kplus~and \kmin~yields by means of model calculations and detector simulations (see reference~\cite{Appelshauser:1998yb}). Here we identify the protons and antiprotons directly by means of their specific energy loss in the MTPCs. 
The new results should therefore be more reliable due to the smaller and better determined corrections leading to smaller systematic uncertainties. The two results are consistent within the systematic errors.

\
\begin{figure}[h]
\includegraphics[width=0.6\linewidth]{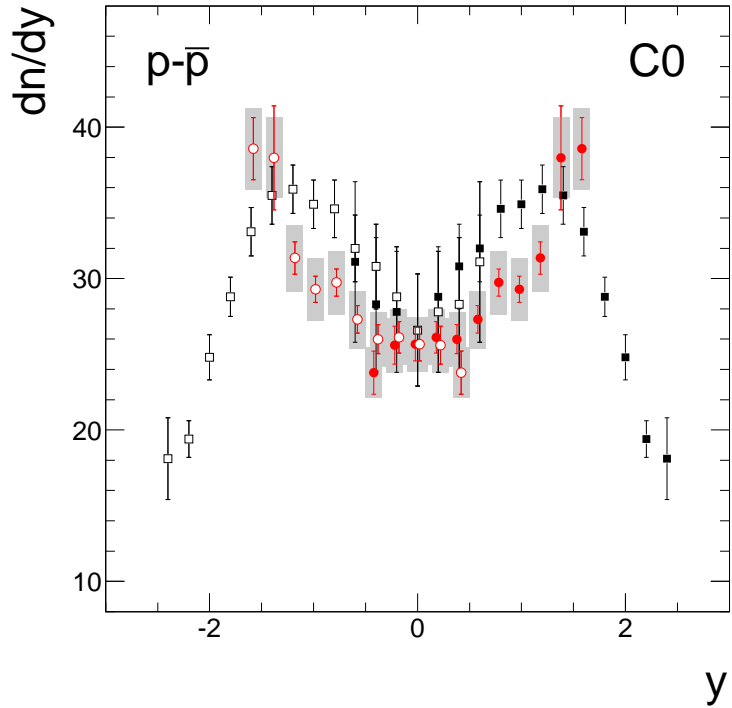}
\caption{The net-proton rapidity distribution at 158\agev~for centrality bin C0 is compared with previously published data {\protect\cite{Appelshauser:1998yb}}. The full circles indicate results from the analysis presented in this paper, whereas the full squares show the previously published data. The open points result from reflection of the data points at midrapidity. The shaded bars represent the sytematic uncertainties. Error bars represent statistical errors. }
\label{fig:C0_net-prot_comp} 
\end{figure}

\subsection{Model comparisons and conclusions}\label{mod_com}

\begin{figure}[!h]
\includegraphics[width=0.7\linewidth]{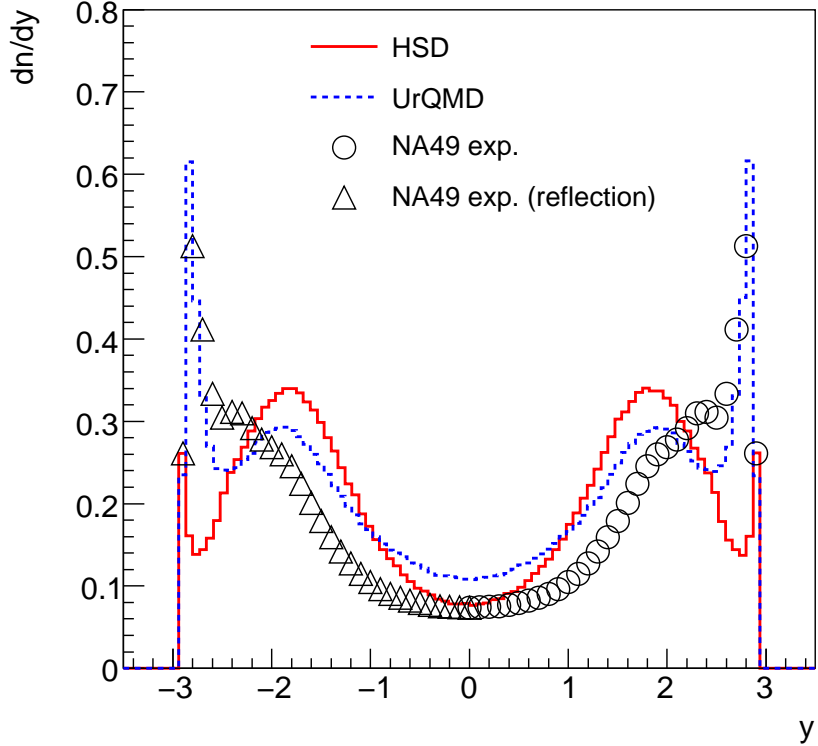}
\caption{Rapidity distribution of net-protons from p+p-collisions at 158~GeV/c beam momentum~{\protect\cite{Fischer}}. The solid line represents the HSD {\protect\cite{Weber:2002qb}} and the dashed line the UrQMD {\protect\cite{Petersen:2008kb}} calculation. }
\label{fig:pp_comp_modell} 
\end{figure} 

\begin{figure}[h]
\includegraphics[width=0.9\linewidth]{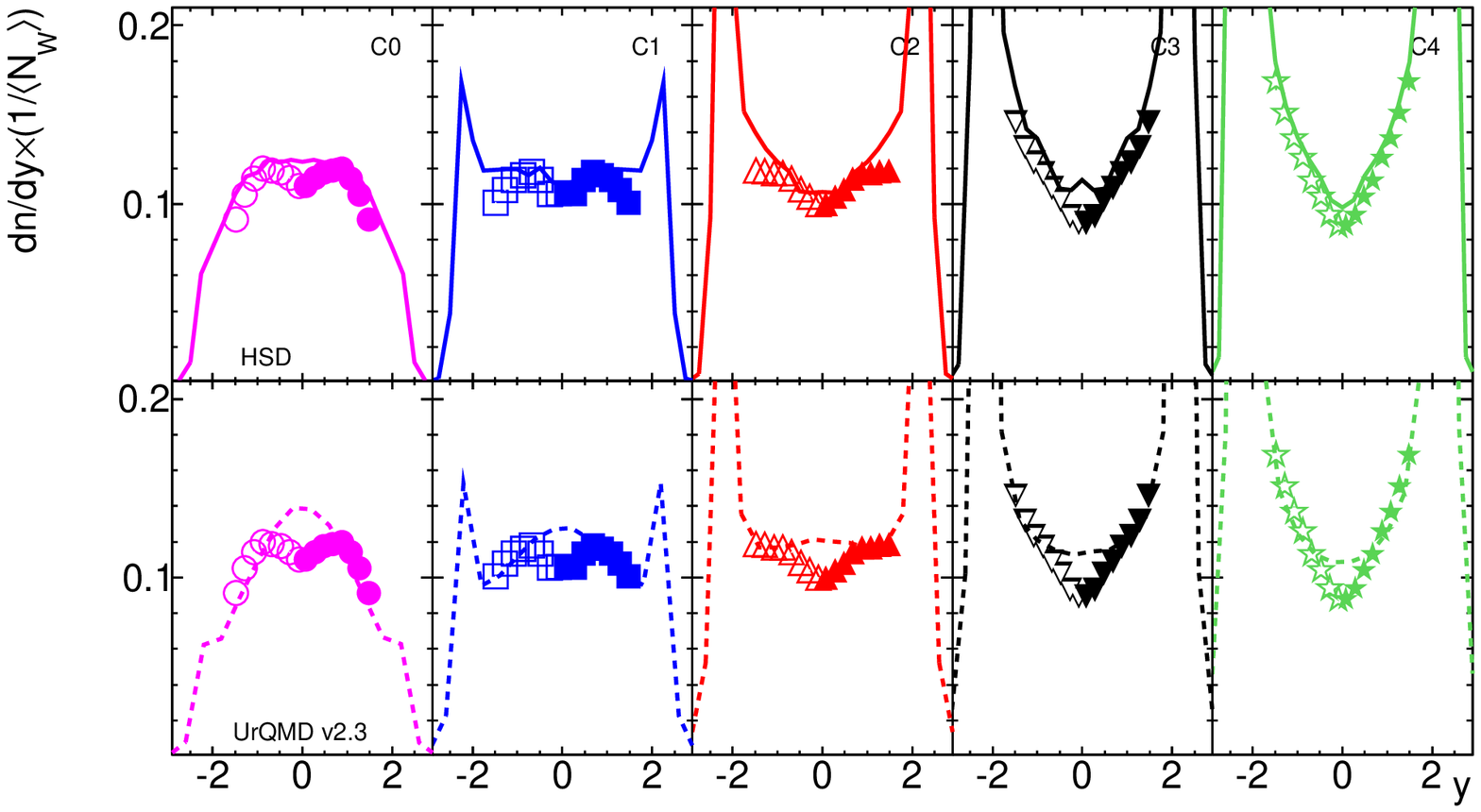}
\caption{The rapidity distributions scaled with 1/\nwound~of 
net-protons for five 
different centralities at 40\agev~are shown together with results of HSD (top) {\protect\cite{Weber:2002qb}} 
and UrQMD (bottom) {\protect\cite{Petersen:2008kb}} calculations. For the statistical errors see \Fi{fig:dndy_p_40_158}. For the systematic errors see section IIIC. }
\label{fig:40_comp_model} 
\end{figure} 

\begin{figure}[h]
\includegraphics[width=0.9\linewidth]{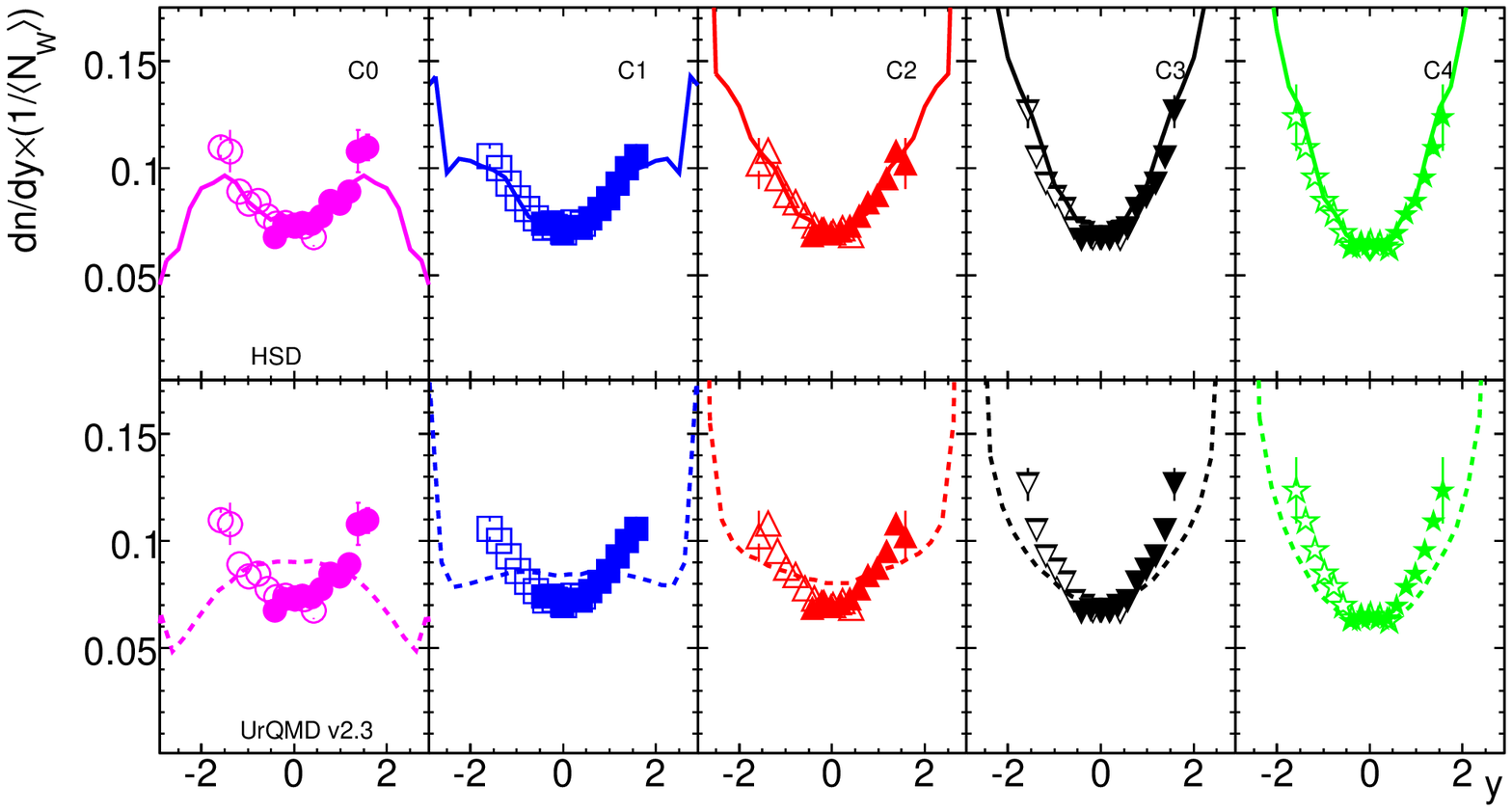}
\caption{The rapidity distributions scaled with 1/\nwound~ of net-protons for five different centralities at 158\agev~ are shown together with results of HSD (top) {\protect\cite{Weber:2002qb}} and UrQMD (bottom) {\protect\cite{Petersen:2008kb}} calculations. For the statistical errors see \Fi{fig:dndy_p_40_158}. For the systematic errors see section IIIC. 
}
\label{fig:158_comp_model} 
\end{figure}

Net-proton spectra are compared with HSD~\cite{Weber:2002qb} 
and UrQMD-2.3~\cite{Petersen:2008kb} model calculations. 
Before addressing possible differences between the model results and 
Pb+Pb collision data, we check how well the models reproduce the rapidity distribution of net-protons in p+p interactions. It is important to note here that the model calculations include also elastic scattering whereas the experimental data represent inelastic interactions only. 
\Fi{fig:pp_comp_modell} reveals significant differences between the models and the experimental data. First, UrQMD~\cite{Petersen:2008kb} has more stopping than HSD~\cite{Weber:2002qb}  with a similar shape as the data, but a significantly higher yield at midrapidity. Second, HSD reproduces the midrapidity yield but fails to reproduce the shape.
These shortcomings render comparison to Pb+Pb data questionable.
The comparison is shown nevertheless in \Fi{fig:40_comp_model} for 
40\agev~and in \Fi{fig:158_comp_model} for 158\agev, in which the rapidity density distributions are divided by the average number of wounded nucleons \nwound. Note that antiprotons were not subtracted in the 40\agev~data, 
since their yield is negligibly small compared to the proton yield ($<~1\%$ at midrapidity).  Overall we observe fair agreement between the HSD model calculation and the data in the rapidity region covered by the NA49 measurements. Larger disagreements are seen for the UrQMD predictions. The differences can be attributed either to known differences between the models (see below) or to the deficiencies in describing proton distributions in elementary p+p interactions. 
The HSD~\cite{Weber:2002qb} calculations describe the data always better than UrQMD~\cite{Petersen:2008kb} calculations. These significant differences in the rapidity distributions between HSD and UrQMD are due to a different definition of "formed" and "unformed" hadrons. In HSD a hadron is considered as "formed" only if the energy density (in the surrounding cell) drops below a critical value, which is taken to be 1~GeV/fm$^{3}$ in line with results from lattice QCD on the critical energy density for deconfinement. Otherwise the hadron is considered as "unformed" and unable to interact with other hadrons until the energy density drops to the critical value. This energy density criterion is not included in UrQMD which leads therefore to a substantial overestimate of the energy loss of participant nucleons in collisions in which high energy densities are reached~\cite{Weber:2002qb}. 

In Pb+Pb collisions at 158\agev~we observe a parabolic shape near to midrapidity ($|y|<$ 1) which is similar to the one seen in inelastic p+p
interactions as well as in HSD calculations. This can be explained by an 
exponential decrease of the proton yield towards midrapidity
(according to exp[-($y-y_{beam})$]), which in turn results from the approximately flat probability density distribution as a function of Feynman x~\cite{Busza}. At 40\agev~the rapidity distributions are similar to those at 158\agev~for more peripheral collisions, while they develop a double hump structure for central collisions. The normalized mid-rapidity yield increases by about 25\% at 40\agev~and 10\% at 158\agev~from the most peripheral to the most central collisions in line with findings in earlier analyses of proton yields at mid-rapidity~\cite{Alt:2005gr} and recent results on hyperons~\cite{na49_hyp_cent}. HSD calculations reproduce this trend quantitatively.

The significant differences between the proton distributions at 
40\agev~and 
158\agev~are at least partly due to the differences in experimental 
acceptances at the two energies. The change in shape which occurs away 
from midrapidity at 40\agev~is probably also present at the higher energy, 
however, in a range $|y|~<~(y_{max}-1.2)$ which is outside the NA49 acceptance.
Finally, the integral of the normalized net proton distributions over the measured 
region increases by 8\% at 40\agev~and by 5\% at 158\agev~with decreasing 
centrality. This is an indication that the spectrum of net protons contains 
nucleons not only from the nuclear overlap region specified in the Glauber model (which we call wounded nucleons) but also from the spectator remnants. Such an additional 
contribution to the proton spectra is confirmed in the results 
from the model calculations in which the coverage extends from target 
to beam rapidity (see Figs.~\ref{fig:40_comp_model} 
and~\ref{fig:158_comp_model}).
The origin of these "participating spectators" could be elastic and inelastic N+N$_{spectator}$ and meson+N$_{spectator}$ interactions (N stands for nucleon). 
Since there is no way to separate the "participating" spectator protons 
from the wounded nucleons in the experimental
data, it is  difficult to draw conclusions on nuclear
stopping from the study of net proton spectral shapes in centrality 
selected Pb+Pb collisions. On the other hand this spectator contribution 
seems to be absent in the hyperon rapidity distributions~\cite{na49_hyp_cent}, thus the e.g. meson+N$_{spectator}$ interactions are not violent enough to produce hyperons.

\section{Summary}
 The NA49 collaboration analysed proton and antiproton spectra in
 40\agev~and 158\agev~Pb+Pb reactions covering the 43.5\% most central collisions. In the transverse mass spectra no strong variation with centrality is discernible, but the average \mt~increases by roughly 20\% from peripheral to central collisions. The rapidity distributions at 
158\agev~have a common concave shape (Fig.~\ref{fig:dndy_p_40_158}b) which however gets shallower with increasing centrality, best seen if normalized to the number of participants as shown in Fig.~\ref{fig:cc_dndy_all_p_ap}. The rapidity distributions at 40\agev~on the contrary exhibit a strong centrality dependence starting with a concave or 'V' shaped structure at large impact parameters, which turns into a symmetric double hump shape for more central collisions. The minimum at the center, which is characteristic of the concave shape, persists at all centralities. Sizeable contributions from non-participants in the rapidity range $|y|~<~(y_{beam}-1.2)$ are observed, which are probably due to secondary interactions of produced particles in spectator matter. Thus rapidity loss analyses of net proton spectra in nucleus-nucleus collisions have to account for this unwanted component of the spectrum. The midrapidity yield normalized to the number of participants varies only by 10\% (5\%) with centrality at 40\agev~(158\agev) reaching its maximum for the most central collisions. As expected it decreases when going from 40\agev~to 158\agev~(by 25\%). These finding may be compared to the corresponding data from AGS experiment E917 extracted from reference~\cite{Back:2000ru} using reference \cite{Dariusz} to compute the number of participants. We find roughly a factor of two higher (normalized) yields at midrapidity with centrality variations of 3\%, 5\%, and 5\% at 6\agev, 8\agev, and 10.8\agev, respectively. 
Antiproton spectra could not be extracted at 40\agev~because of lack of statistics. At 158\agev~the normalized total (see Fig.~\ref{fig:tot_mult_pbar_2}) and midrapidity yields increase with impact parameter. Such a behavior is expected in case absorption plays a significant role or, equivalently, the baryon rich collision system approaches chemical equilibrium~\cite{Becattini:2005xt}.

We chose to compare our findings to the transport model calculations from the HSD and UrQMD codes. Although both models reproduce rapidity density distributions in p+p collisions only   
with significant deviations as shown in Fig.~\ref{fig:pp_comp_modell}, HSD gives a good description of the nuclear collision data at both energies (see upper rows in Figs.~\ref{fig:40_comp_model} and \ref{fig:158_comp_model}) in the region covered by NA49 measurements with deviations only at large rapidities. The UrQMD model calculations also fit the experimental distributions fairly well, although with larger discrepancies which are most pronounced at midrapidity and in central collisions.

\begin{table}[h]
	\centering
\caption{Proton rapidity densities \dndy~at five different centralities and 40\agev. The quoted errors are statistical. For the systematic errors see section IIIC.}
\vspace{\baselineskip}
\begin{tabular} {|c||c|c|c|c|c|}
\hline
    & \dndy (C0) & \dndy (C1)& \dndy (C2)& \dndy (C3)& \dndy (C4)\\
\hline\hline
-0.02 $\leq y \leq$ 0.18& 38.64$\pm$1.11& 30.69$\pm$0.79&21.01$\pm$0.51& 12.69$\pm$0.33& 8.19$\pm$0.20\\
0.18 $\leq y \leq$ 0.38&40.15$\pm$1.07&30.63$\pm$0.75&21.86$\pm$0.51& 13.26$\pm$0.34&8.74$\pm$0.22\\
0.38 $\leq y \leq$ 0.58&41.23$\pm$1.08&32.93$\pm$0.70&22.72$\pm$0.51& 14.49$\pm$0.34&9.68$\pm$0.23\\
0.58 $\leq y \leq$ 0.78&41.72$\pm$0.98&34.24$\pm$0.74&24.01$\pm$0.52& 15.48$\pm$0.35&10.51$\pm$0.23\\
0.78 $\leq y \leq$ 0.98&42.08$\pm$1.32&33.65$\pm$0.72&24.57$\pm$0.49& 16.64$\pm$0.37&11.71$\pm$0.24\\
0.98$\leq y \leq$ 1.18&40.11$\pm$1.19&32.91$\pm$0.85&24.58$\pm$0.50& 17.48$\pm$0.38&12.71$\pm$0.28\\
1.18 $\leq y \leq$ 1.38&36.64$\pm$0.98&31.29$\pm$0.73&24.84$\pm$0.53& 18.70$\pm$0.42&14.06$\pm$0.33\\
1.38 $\leq y \leq$ 1.58&31.63$\pm$1.09&29.16$\pm$0.76&24.94$\pm$0.56& 20.69$\pm$0.51& 15.69$\pm$0.48\\
\hline
\end{tabular}
\label{tab:prot_40}
\end{table}

\begin{table}[h]
	\centering
\caption{Proton rapidity densities \dndy~at five different centralities and 158\agev. The quoted errors are statistical. For the systematic errors see section IIIC.}
\vspace{\baselineskip}
\begin{tabular} {|c||c|c|c|c|c|}
\hline
    & \dndy (C0) & \dndy (C1)& \dndy (C2)& \dndy (C3)& \dndy (C4)\\
\hline\hline
-0.52 $\leq y \leq$ -0.32& 25.63$\pm$1.41 & 22.17$\pm$0.74  & 14.36$\pm$0.43  &   9.28$\pm$0.30 &   5.84$\pm$0.21\\
-0.32 $\leq y \leq$ -0.12&  27.49$\pm$1.24 & 22.32$\pm$0.61&  15.14$\pm$0.38  &  9.46$\pm$0.25  &  6.03$\pm$0.18\\
-0.12 $\leq y \leq$ 0.08& 27.51$\pm$1.09 & 20.97$\pm$0.52 &  14.79$\pm$0.33  &   9.46$\pm$0.23  &  6.07$\pm$0.19\\
0.08 $\leq y \leq$ 0.28& 28.07$\pm$1.01 &  21.81$\pm$0.51 &  14.96$\pm$0.31  &  9.34$\pm$0.21   & 5.98$\pm$0.16\\
0.28 $\leq y \leq$ 0.48   & 27.93$\pm$0.94 & 21.74$\pm$0.52  & 15.35$\pm$0.30   & 9.64$\pm$0.21 &   5.96$\pm$0.15\\
0.48 $\leq y \leq$ 0.68& 29.20$\pm$0.89 & 22.91$\pm$0.50 &  16.27$\pm$0.30  &  9.92$\pm$0.23  &  6.39$\pm$0.16\\
0.68 $\leq y \leq$ 0.88&31.30$\pm$0.88 & 23.96$\pm$0.49 & 17.4$\pm$0.31 &  10.97$\pm$0.23   &  7.05$\pm$0.17\\
0.88 $\leq y \leq$ 1.08&30.51$\pm$0.85 & 25.10$\pm$0.49 &  17.87$\pm$0.32 &  11.53$\pm$0.24 &    7.55$\pm$0.19\\
1.08 $\leq y \leq$ 1.28&32.35$\pm$1.05  & 26.88$\pm$0.57 &  19.22$\pm$0.39 &  12.25$\pm$0.26 &    8.43$\pm$0.20\\
1.28 $\leq y \leq$ 1.48&38.84$\pm$3.43  & 28.82$\pm$0.58  & 21.46$\pm$0.43 &  13.63$\pm$0.28  &   9.52$\pm$0.22\\
1.48 $\leq y \leq$ 1.68&339.35$\pm$2.00  & 30.30$\pm$1.55 &   20.13$\pm$2.30  &  16.38$\pm$0.98  &  10.73$\pm$1.31\\
\hline
\end{tabular}
\label{tab:prot_158}
\end{table}
\newpage
\begin{table}[h]
	\centering
\caption{Antiproton rapidity densities \dndy~at five different centralities and 158\agev. The quoted errors are statistical. For the systematic errors see section IIIC.}
\vspace{\baselineskip}
\begin{tabular} {|c||c|c|c|c|c|}
\hline
    & \dndy (C0) & \dndy (C1)& \dndy (C2)& \dndy (C3)& \dndy (C4)\\
\hline\hline
-0.52 $\leq y \leq$ -0.12& 1.61$\pm$0.13 &  1.37$\pm$0.09 &  0.92$\pm$0.06 &  0.78$\pm$0.05  & 0.51$\pm$0.04\\
-0.12 $\leq y \leq$ 0.28& 1.75$\pm$0.12 & 1.37$\pm$0.08 &  1.06$\pm$0.05 &   0.80$\pm$0.04  & 0.55$\pm$0.03\\
0.28 $\leq y \leq$ 0.68& 1.76$\pm$0.11 & 1.43$\pm$0.07 &  0.96$\pm$0.05 &  0.70$\pm$0.04  & 0.49$\pm$0.03\\
0.68 $\leq y \leq$ 1.08& 1.31$\pm$0.10 & 1.11$\pm$0.07 &  0.80$\pm$0.05 &  0.54$\pm$0.038 &  0.36$\pm$0.03\\
1.08 $\leq y \leq$ 1.48& 0.60$\pm$0.10 & 0.57$\pm$0.09 & 0.37$\pm$0.05 &  0.21$\pm$0.03  & 0.31$\pm$0.06\\
\hline
\end{tabular}
\label{tab:aprot_158}
\end{table}

\begin{acknowledgments}
Acknowledgements: This work was supported by
the US Department of Energy Grant DE-FG03-97ER41020/A000,
the Bundesministerium fur Bildung und Forschung, Germany (06F 137),
the Virtual Institute VI-146 of Helmholtz Gemeinschaft, Germany,
the Polish Ministry of Science and Higher Education (1 P03B 006 30, 1 P03B 127 30, 0297/B/H03/2007/33, N N202 078735,  N N202 204638),
the Hungarian Scientific Research Foundation (T032648, T032293, T043514),
the Hungarian National Science Foundation, OTKA, (F034707),
the Bulgarian National Science Fund (Ph-09/05),
the Croatian Ministry of Science, Education and Sport (Project 098-0982887-2878)
and
Stichting FOM, the Netherlands.
\end{acknowledgments}



\end{document}